# Detection and characterisation of icy cavities on the nucleus of comet 67P/Churyumov–Gerasimenko

Philippe Lamy,[1]* Guillaume Faury,[2] David Romeuf,[3] Olivier Groussin[4]

[1]*Laboratoire Atmosphères, Milieux et Observations Spatiales, CNRS & UVSQ, 11 boulevard d'Alembert, 78280 Guyancourt, France*
[2]*Institut de Recherche en Astrophysique et Planétologie, 14 avenue Edouard Belin, 31400 Toulouse, France*
[3]*Université Claude Bernard Lyon 1, 43 boulevard du 11 novembre 1918, 69622 Villeurbanne, France*
[4]*Aix Marseille Université, CNRS, CNES, LAM, 38 rue Frédéric Joliot-Curie, 13388 Marseille cedex 13, France*

## ABSTRACT

We report on the detection of three icy cavities on the nucleus of comet 67P/Churyumov-Gerasimenko. They were identified on high-resolution anaglyphs built from images acquired by the OSIRIS instrument aboard the Rosetta spacecraft on 2016 April 9-10. Visually, they appear as bright patches of typically 15 to 30 m across whose large reflectances and spectral slopes in the visible substantiate the presence of sub-surface water ice. Using a new high-resolution photogrammetric shape model, we determined the three-dimensional shape of these cavities whose depth ranges from 20 to 47 m. Spectral slopes were interpreted with models combining water ice and refractory dark material and the water ice abundances in the cavities were found to amount to a few per cent. The determination of the lifetime of the icy cavities was strongly biased by the availability of appropriate and favourable observations, but we found evidences of values of up to two years. The icy cavities were found to be connected to jets well documented in past studies. A thermal model allowed us to track their solar insolation over a large part of the orbit of the comet and a transitory bright jet on 2015 July 18 was unambiguously linked to the brief illumination of the icy bottom of one of the cavities. These cavities are likely to be the first potential subsurface access points detected on a cometary nucleus and their lifetimes suggest that they reveal pristine sub-surface icy layers or pockets rather than recently recondensed water vapor.

## 1 INTRODUCTION

Potential subsurface access points (SAPs) defined as openings on the surface visible by remote sensing of planetary bodies are ubiquitous in the solar system. According to Wynne et al. (2022b), 3,545 SAPs in the form of pits, vents, fissures, and caves have been identified on 11 bodies and are attracting considerable attention as they offer an access to near surface geology and to preserved volatiles in the form of ices without the need for drilling. The inclusion of four contributed white papers in the 2023–2032 Decadal Survey covering different aspects of the investigation of these features, prominently planetary caves, is particularly revealing.

Space observations of cometary nuclei have revealed a wealth of complex geomorphological features (e.g. Thomas et al. (2015)), but none could be unambiguously considered as SAPs according to Wynne et al. (2022a). In the case of the *Rosetta* mission to comet 67P/Churyumov-Gerasimenko (hereafter comet 67P), 18 quasi-circular pits with diameters ranging from 50 to 300 m have been extensively characterized by Vincent et al. (2015) who found that their formation is likely sublimation driven and probably created by a sinkhole process (Mousis et al. 2015). This led Wynne et al. (2022a) to surmise that their speleogenic nature remains speculative. Likewise, the numerous fractures several hundred meters long present on the nucleus of 67P (El-Maarry et al. 2015) whose depth could not be estimated cannot be considered as SAPs. In summary, even with the large set of high resolution images produced by the *Rosetta* mission, we did not have yet unambiguous evidences of the presence of SAPs on a cometary nucleus.

In contrast, bright features are ubiquitous on the surface of cometary nuclei and are often interpreted as ice patches. The first detection was reported in the case of comet 9P/Tempel 1 by Sunshine et al. (2006) and the presence of solid water ice was ascertained on the basis of their anomalous color with respect to that of the overall nucleus and the presence of water ice absorption bands at 1.5 and 2.0 μm. These authors further showed that a mixture of 3 to 6 per cent water ice grains mixed with nearby dark material matched well the observed absorptions, but remarked that the ice may have recondensed from recent activity. The *Rosetta* mission confirmed the widespread presence of such features and Pommerol et al. (2015) reported the detection of over one hundred meter-sized bright spots in numerous regions of different geomorphologies, but preferably in areas of low insolation. In addition to their brightness, their spectral reflectance over the 300–1000 nm range were found to have red slopes less steep than that of the nearby dark terrains, henceforth a relative blue color in comparison with those terrains. This led to the designation of these features as blue patches abbreviated to BPs that could stand as well for bright patches.

Barucci et al. (2016) selected 13 BPs and confirmed the presence of water ice in eight of them based on infrared spectra, an identification also implemented by Filacchione et al. (2016) on two debris falls in the Imhotep region of the nucleus. The former authors noted that the number of the BPs reached a peak during the perihelion passage of comet 67/P in August 2015. Deshapriya et al. (2016) investigated four BPs in the Khonsu region that have been exposed to sunlight after a long winter: one of them was found to have persisted over more than five months while a small one lasted only a few days. This led Oklay et al. (2017) to investigate the long-term sustainability of water ice patches and these authors found that large clusters are stable over typically 0.5 yr, shrinking around perihelion while small patches disappeared. But even more surprising was the detection of patches surviving up to 2 yr, suggesting their existence during the cometâĂŹs previous orbit. Subsequently, large numbers of BPs were cataloged to study their ensemble properties, their evolution, and their connections to jet activity. Deshapriya et al. (2018) considered a set of 57 BPs which was extended to over 600 by Fornasier et al. (2023) who analyzed them in a homogeneous way. We highlight two of their main conclusions: i) the cumulative area of the BPs remains very small, typically 0.1 per cent of the total surface of the nucleus and ii) the bulk of the BPs have an area less than 1 m$^2$. The latter conclusion supports the scenario of Ciarniello et al. (2022) in which water ice enriched blocks (WEBs) of 0.5–1 m size are embedded in a refractory matrix. The progressive erosion of this matrix results in the emergence of these WEBs and their exposure to solar illumination makes them observed at BPs possibly becoming sources of jets.

Cavities (as well as alcoves) on the nucleus of 67P have often been advocated to explain the presence of BPs as the shadows cast by these structures facilitate the recondensation and redeposition of sublimating ices during cometary nights, a process demonstrated by laboratory experiments (Sears et al. 1999). In their catalogue of more than 600 BPs, Fornasier et al. (2023) associated 13 of them to cavities as displayed in their map of the nucleus (their fig. 1). They were further identified as the source of jets by Fornasier et al. (2019) who displayed a similar map (their fig. 1). However, no details on these cavities were provided and their very existence was inferred from the local topographies suggested by the images. The only extensive study of cavities remains that of Hasselmann et al. (2019) who showed that four out of the six considered were recently formed over a time interval of one year and a half, a probable consequence of the strong insolation of the southern Khonsu region during the passage of the comet at perihelion. Concentrating on the three cavities that harbor icy patches, their size ranges from 16 to 63 m and their depth estimated from the length of the projected shadows, from 1.4 to 2.8 m, hence very shallow (we interpreted "Eff. height" in their table 3 as depth). Although these cavities can hardly be considered as subsurface access points, these authors made the interesting remark on the "Circumstantial evidence through spectrophotometric analysis and also source/outburst mass comparison that likely indicates the presence of ice-rich pockets at tens of meters underneath" the cavities.

In this article, we report on the unambiguous detection and on the characterization of three icy cavities so defined by their bright bottom presumed to result from the presence of ice. These narrow openings, 20 to 30 m across, offer a glimpse inside the nucleus at depths of several ten meters and make them truly subsurface access points. After briefly summarizing the *Rosetta* mission and the OSIRIS/NAC instrument, Section 2 describes the observations and the images and stereo anaglyphs used in the analysis. The detection, localization and regional settings of the cavities are the subjects of Section 3 followed by a presentation of their geometric (Section 4), photometric and spectrophotometric (Section 5) properties. The derivation of the abundance of water ice is presented in Section 6. Section 7 considers the questions of the lifetime of the icy cavities and their connections to cometary jets. We investigate and quantify the thermal constraints on the cavities in Section 8. We discuss our results in Section 9 and finally conclude in Section 10.

## 2 OBSERVATIONS AND METHODS

The international *Rosetta* mission was conceived as a rendezvous with comet 67P/Churyumov-Gerasimenko while inactive at a large heliocentric distance so as to study its nucleus, followed by an escort phase and past perihelion to characterize the development of its activity. After its launch on 2004 March 2, the spacecraft arrived at a distance of 100 km from the nucleus on 2014 August 6 and then started a complex journey around the nucleus in order to fulfill its scientific mission which ended on 2016 September 30. Various constraints besides scientific dictated this circum-cometary navigation, particularly the safety of the spacecraft. Consequently, its distance to the nucleus varied considerably as illustrated in Fig. 1 and conspicuously increased around perihelion time (2015 August 13) as a measure of protection against the increasing cometary activity. The variations of the spatial resolution as well those of the illumination and viewing directions imposed major challenges to the analysis and interpretation of the images of the nucleus, especially when investigating small features and even more, their depth. It should therefore be realized that the inventory of BPs and their lifetime is seriously hampered and biased by the observing conditions.

The present analysis exclusively relies on the images obtained with the Narrow Angle Camera (NAC) of the Optical, Spectroscopic, and Infrared Remote Imaging System (OSIRIS) as described by Keller et al. (2007). Briefly, the NAC was a three-mirror anastigmat equipped with a 2048 × 2048 pixel backside illuminated CCD detector with a UV optimized anti-reflection coating, and a double filter wheel holding a total of 13 filters. Images used in the present study were obtained with the following three filters: F82, F84, F88 whose properties are listed in Table 1. The pixel size of 13.5 µm yielded an image scale of 3.9 arcsec pixel$^{-1}$.

The NAC acquired approximately 27,000 images during the whole mission and their fully corrected and calibrated versions are available on ESA's Planetary Science Archive. For each image, the data are organized in nine layers, the image itself, the distance per pixel from the camera to the target surface, the emission, incidence, and phase angles for each pixel, the facet identification related to the shape model of the nucleus described below, and the $x$, $y$, $z$ coordinates of each pixel in the body-fixed frame. The reference three-dimensional model "SHAP7" elaborated by Preusker et al. (2017) has about 44 million facets (1–1.5 m horizontal sampling). The body-fixed frame has its origin at the center of gravity of the nucleus, its z-axis along its rotational axis and the definition of zero longitude (x-axis orientation) is such that the boulder-like Cheops in the Imhotep region has a longitude of 142.3°(right-hand-rule eastern longitude) and a latitude of âĂŞ0.28°forming the so-called Cheops reference system of coordinates.

Beside images, the present work extensively relies on the use of anaglyphs whose merits for scientific analysis are now well recognized. For instance, the HiRISE team[1] has placed a major emphasis on stereo imaging needed to make small-scale topographic measurements, essential to the characterization of candidate landing sites and to the quantitative study of surface processes. Their anaglyphs are presented on equal terms with digital terrain models on their website, thus emphasizing their complementary nature. In the case of 67P, several published articles have already included anaglyphs to support their scientific analysis, for instance Auger et al. (2015), Mottola et al. (2015), and Pajola et al. (2017). Our team has undertaken the systematic production of anaglyphs of nucleus and coma of 67P (Lamy et al. 2019); together with their documented presentation, they form a catalogue publicly accessible on a dedicated website[2], hosted by the Centre National d'Etudes Spatiales (CNES). Our anaglyphs implement the standard red/cyan system: the left image is coded in red levels and the right image is code in cyan (green+blue) levels. Instructions for optimal viewing are available on the above website. At time of writing, 1,823 anaglyphs have been produced after analyzing a subset of approximately 13,000 NAC images over a time interval extending from 2014 August 3 to 2016 September 30. The present article incorporates a restricted number of anaglyphs of the cavities. Many more may be found on the above website by selecting the tab "Features" and then "Pits".

---

[1] https://hirise.lpl.arizona.edu/

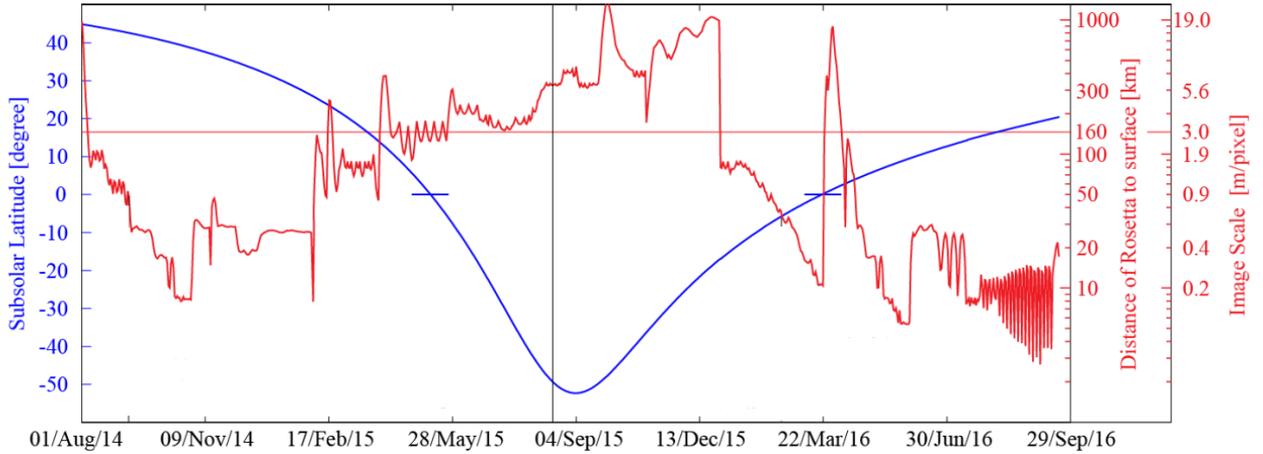

**Figure 1.** Timeline of observation and illumination conditions during the *Rosetta* campaign at comet 67P: subsolar latitude (blue curve), distance between the spacecraft and the surface of the nucleus and corresponding image scale of the NAC images (red curve). The grey vertical line indicates the perihelion passage on 2015 August 13. This is a simplified reproduction of fig. 2 of Preusker et al. (2017).

**Table 1.** Spectral properties of the NAC filters and solar irradiance at 1 AU. $\lambda_c$ and $\Delta\lambda$ are the central wavelength and the bandwith of the filters, respectively, both expressed in units of nm. The solar irradiance is expressed in units of $\mathrm{W\,m^{-2}\,nm^{-1}}$

| Code | Name | $\lambda_c$ | $\Delta\lambda$ | $F_\odot$ |
|------|------|------|------|------|
| F82 | Orange | 649.2 | 84.5 | 1.98 |
| F84 | Blue | 480.7 | 74.9 | 1.59 |
| F88 | Red | 743.7 | 64.1 | 1.30 |

## 3 DETECTION AND LOCALIZATION OF THE CAVITIES

The three deep icy cavities $C_A$, $C_B$, and $C_C$ were discovered serendipitously during the routine process of visual inspection and validation of the anaglyphs before their incorporation in the dedicated website. Their size of less than 35 m and their rather large depth-to-diameter ratio $d/D$, particularly for two of them, required very favorable conditions to detect their icy bottom in terms of both spatial resolution and geometric conditions (illumination and viewing directions). In fact, this may explain why very few were detected and we surmise that many more were probably present on the nucleus of 67P.

These favorable conditions were satisfied during the low-altitude flyby of the nucleus that took place from 2016 April 9 at ≈23 h to April 10 at ≈1 h, almost eight months after perihelion passage, the comet being at a heliocentric distance ranging from 2.76 to 2.78 AU. Over this time interval of ≈2 h, the mean distance between the spacecraft and the nucleus surface varied between 28.6 and 29.2 km, the phase angle between 1° and 8°, and the NAC pixel scale between 0.53 and 0.54 m. The NAC images and resulting anaglyphs of this observational sequence constituted the primary dataset of our investigation. Additional, complementary images from other sequences were further included in support of our analysis as appropriate.

The three cavities were found on the outer part of the big lobe of the nucleus as displayed in Figure 2 on i) a large scale image of March 10, 2015 and ii) on a regional map reproduced from Thomas et al. (2018). $C_A$ and $C_B$ are located in the Kephry region close to the border with the Imhotep region and $C_C$ in the Ash region, very close to the border with the Kephry region. Both regions are characterized by consolidated material in the form of terraces and layers (well depicted in fig. 4 of Massironi et al. (2015) and fig. 5 of Lee et al. (2016)) as well as smooth deposits.

Considering the sub-regions introduced by Thomas et al. (2018), $C_A$ and $C_B$ belong to Kephry-c, a highly complex unit with rough, rocky terrain, smoother coatings, and boulders. $C_C$ belongs to Ash-i characterized by a large-scale rough terrain especially at the boundaries with the Imhotep and Kephry regions. A finer description is offered by Feller et al. (2019) who used the same observational sequence of 2016 April 9 and 10 to precisely study the so-called Imhotep–Khepry transition. Their fig. 2 displays the geomorphological mapping of this transition area and indicates that $C_A$ and $C_B$ reside in an outcropping stratified terrain whereas $C_C$ is on a talus above a gravitational accumulation of debris.

Our anaglyphs go one step further and offer impressive views of the sites of the three cavities with three-dimensional renderings that allow understanding their geomorphology. Figures 3, 4, and 5 are devoted to cavities $C_A$, $C_B$, and $C_C$, respectively. Each figure displays three anaglyphs at low (for an overview), medium, and high spatial resolutions.

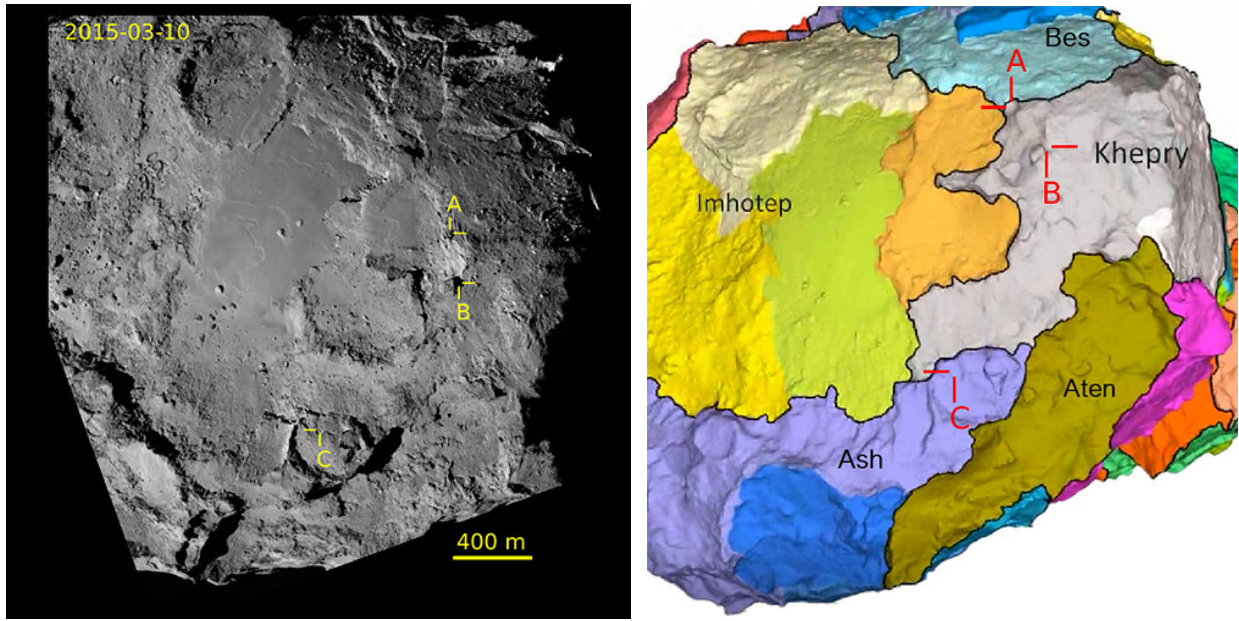

**Figure 2.** Location of the three cavities on a large scale image of the big lobe (left panel) and a regional map of Thomas et al. (2018). Within each region bounded by black lines, sub-regions are characterized by different colours.

$C$A is located in the highly degraded section of the scarp that partly surrounds the southern depression adjacent to the border with Imhotep. The debris of this degradation are clearly seen as an accumulation of small boulders and coarse materials extending from the scarp to about half of the depression. The asymmetric funnel-shaped opening of $C$A narrows down with increasing depth forming and approximately circular tube to the bright icy patch at the bottom.

$C$B is located very near the top of the same scarp as $C$A but closer to the cuesta that separates the two twin depressions, the two cavities being separated by 247 m. The local geomorphology appears highly complex with, in particular, a remarkable mound, $50 \times 85$ m in size, overlooking its view during the early phase of the observational sequence; fortunately the viewing conditions evolved and $C$B became fully visible later on. The mound looks distinctly darker than the surrounding terrain and we consequently nicknamed it "Kemour", the Egyptian name for "the great black", one of the numerous manifestations of Osiris. Kemour is likely the isolated remnant of a higher layer that has otherwise totally disappeared in the past, its darkness resulting from its longer exposure to solar illumination and thus, prolonged irradiation of its organic content. As such, Kemour would be best described as a witness butte or outlier, two relevant geological terminologies.

$C$C is located approximately 90 m from the scarp that marks the limit between the Ash and Kephry regions, precisely in a degraded talus that apparently results from the partial collapse of the surrounding large depression. Accumulation of coarse material is conspicuous at the foot of the talus. As we shall see later, $C$C is shallower than the two other cavities and we suspect that it may be partly filled with debris, hence an inhomogeneous icy bottom with the brightest part restricted to only one side of the cavity. This filling looks even more pronounced in the case of the neighboring smaller cavity $C$C' located approximately 60 m away from $C$C with only a few bright points. Consequently, its characterization will only be briefly addressed in the context of the three-dimensional reconstruction of the cavities.

Pommerol et al. (2015) found that isolated BPs can be observed in all types of regions so that "the relation between the BPs and the surrounding terrain is generally unclear". This is confirmed by Fornasier et al. (2023) who pointed out that they "are observed in different types of morphological terrains". The same situation appears to prevail for our cavities according to our above description of the different surrounding terrains although with only three of them, it is difficult to ascertain a global trend. Vincent et al. (2015) did not address this question in the case of pits, but their figure 1 indicates that they were found in very different regions.

The accurate geo-location of the centre of each cavity is given in Table 2 in Cartesian and geographic coordinates using the reference system of Preusker et al. (2017) introduced in Section 2.

## 4 GEOMETRIC PROPERTIES OF THE CAVITIES

We first performed a two-dimensional characterization of the three cavities on selected images of the April sequence that offer optimal viewings of their bright bottom (i.e. maximizing their area) and choosing those taken with the F84 filter offering the best contrast with respect to the local terrains. Table 3 gives two characteristic dimensions as measured along the red lines shown in Figure 6 and the area calculated after applying a mask on each image. Each mask was constructed from the image itself by applying a threshold that allowed isolating the bright region. Considering the equivalent circular area, we then derived an effective diameter. It amounts to 15.2 and 19.6 m for $C$A and $C$B, respectively, but appears significantly larger, 32.0 m, for $C$C. These sizes are significantly smaller than those of the pits studied by Vincent et al. (2015) whose diameters range from 50 to 300 m and significantly larger than those of the isolated BPs, about one to two meters (Pommerol et al. 2015), consistent with the result of Fornasier et al. (2023) based on more than 600 volatile exposures, typically a few square meters or smaller.

The spatial resolution of the SHAP7 model is insufficient to achieve a detailed three-dimensional characterization of the three cavities. We therefore carried out our own reconstruction that will be described in a forthcoming article. Briefly, it makes use of all im-ages taken after perihelion from 2015 October 26 to 2016 February 10, that is 7,682 NAC and 1,504 WAC images and performs a photogrammetric processing. The resulting high resolution 3D model "67P-133M" of the whole nucleus has 133 million facets (66 million vertices) to be compared with about 44 million facets for the SHAP7 model. Figure 7 illustrates the dramatic difference between the two models in the case of cavity $C_A$ as it is literally smoothed out on the SHAP7 model. Figure 8 presents an image and the associated anaglyph of a region of the nucleus encompassing the three cavities constructed from our model. A texture derived from the images themselves was applied to the shape model for a realistic rendering of the nucleus surface as described in Appendix A.

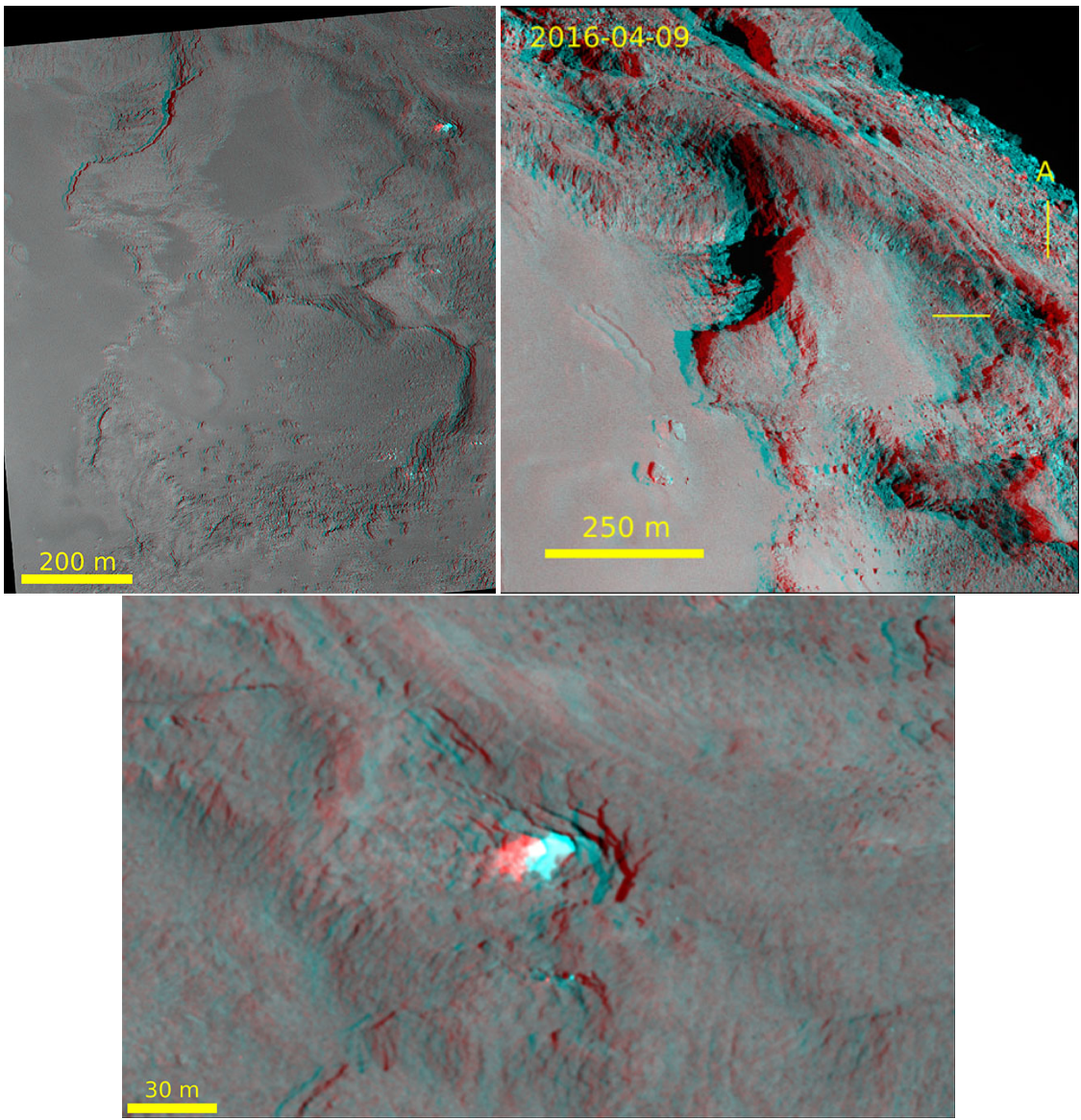

**Figure 3.** Three anaglyphs of cavity $C_A$ at low (upper left panel), medium (upper right panel), and high (lower panel) spatial resolutions. The hypertext links connect to the original anaglyphs of the on-line catalogue.

**Table 2.** Cartesian and geographic coordinates of the centre of the cavities.

| Cavity | x ; y ; z (m) | Lat (deg) | Lon (deg) |
|--------|---------------|-----------|-----------|
| $C_A$  | -251 ; 1479 ; -803 | -28.15 | 99.62 |
| $C_B$  | -195 ; 1654 ; -623 | -20.52 | 96.71 |
| $C_C$  | -554 ; 1476 ; 376  | 13.41  | 110.58 |

**Table 3.** Properties of the three cavities: size, area, effective diameter ($D$), depth ($d$), and depth-to-diameter ratio ($d/D$).

| Cavity | Size (m) | Area (m$^2$) | $D$ (m) | $d$ (m) | $d/D$ |
|--------|----------|--------------|---------|---------|-------|
| $C_A$  | 18.4 × 16.2 | 181 | 15.2 | 47 | 3.1 |
| $C_B$  | 24.2 × 18.9 | 303 | 19.6 | 20 | 1.0 |
| $C_C$  | 35.0 × 32.5 | 804 | 32.0 | 20 | 0.6 |

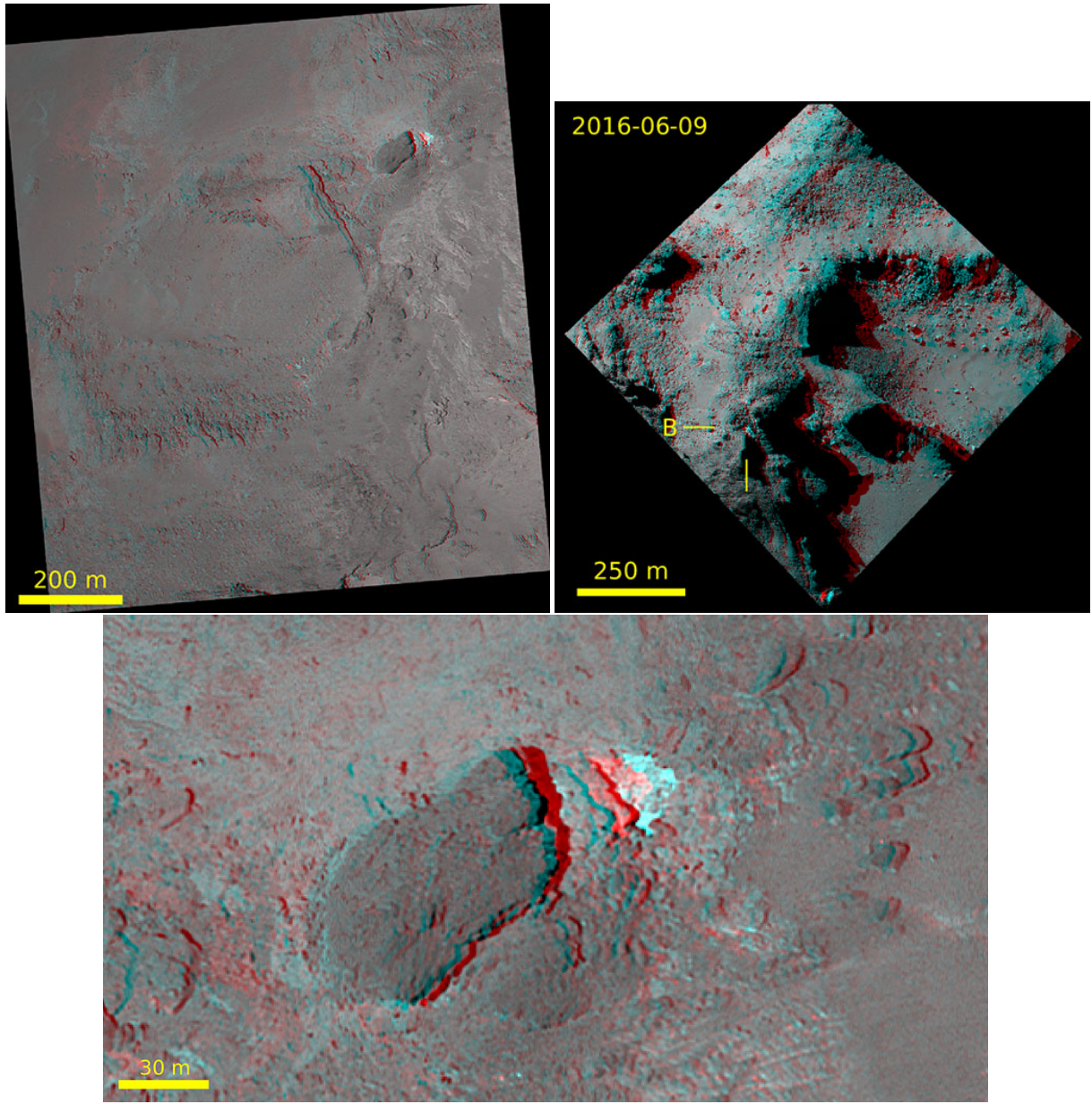

**Figure 4.** Three anaglyphs of cavity $C_B$ at low (upper left panel), medium (upper right panel), and high (lower panel) spatial resolutions.

Local DEMs of the $C_A$, $C_B$, and $C_C$ cavities were extracted by defining parallelepiped boxes encompassing each one. The boxes were visually oriented by imposing that each top plane was parallel to the area surrounding its cavity, the perpendicular direction being approximately the local vertical. Two profiles at right angles were extracted along the longest (LL) and shortest (SS) dimensions of the cavities and going through the deepest points as displayed in Figures 9, 10, and 11. Additional profiles are displayed in three cases: i) for $C_B$, an enlarged transverse profile S'S' including the mound, ii) for $C_C$, a diagonal profile DD, and iii) for $C_{C'}$, a transverse profile S'S', all through the respective deepest points. The most interesting result is obviously their depth which is not always easy to specify owing to the rugged surrounding terrains. When appropriate, we basically drew on each profile a straight line joining the two external rims of a given cavity as a realistic reconstruction of the terrain in the absence of the cavity and measured its vertical distance to the deepest point of the profile. For $C_A$, the two profiles yield consistent depths of 48 m (LL) and 46 m (SS), thus an average of 47 m. For $C_B$, the LL profile is useless for our present purpose, but the SS and S'S' profiles indicate the same depth of 20 m. For $C_C$ and $C_{C'}$, we favored the LL profile as the others are biased by the overall slope of the depression; the resulting depths amount to 20 and 14 m, respectively. Table 3 summarizes the properties of the three cavities including the depth-to-diameter ratio $d/D$ which ranges from 0.6 ($C$C) to 3.1 ($C$A), with an intermediate value of 1.0 for $C$B. As a matter of comparison, the large pits (diameter >= 50 m) studied by Vincent et al. (2015) have values of $d/D$ ranging from 0.45 to 0.9 for the active ones and from 0.1 to 0.5 for the then inactive ones. Clearly $C$B and above all $C$A deserve their classification as cavities.

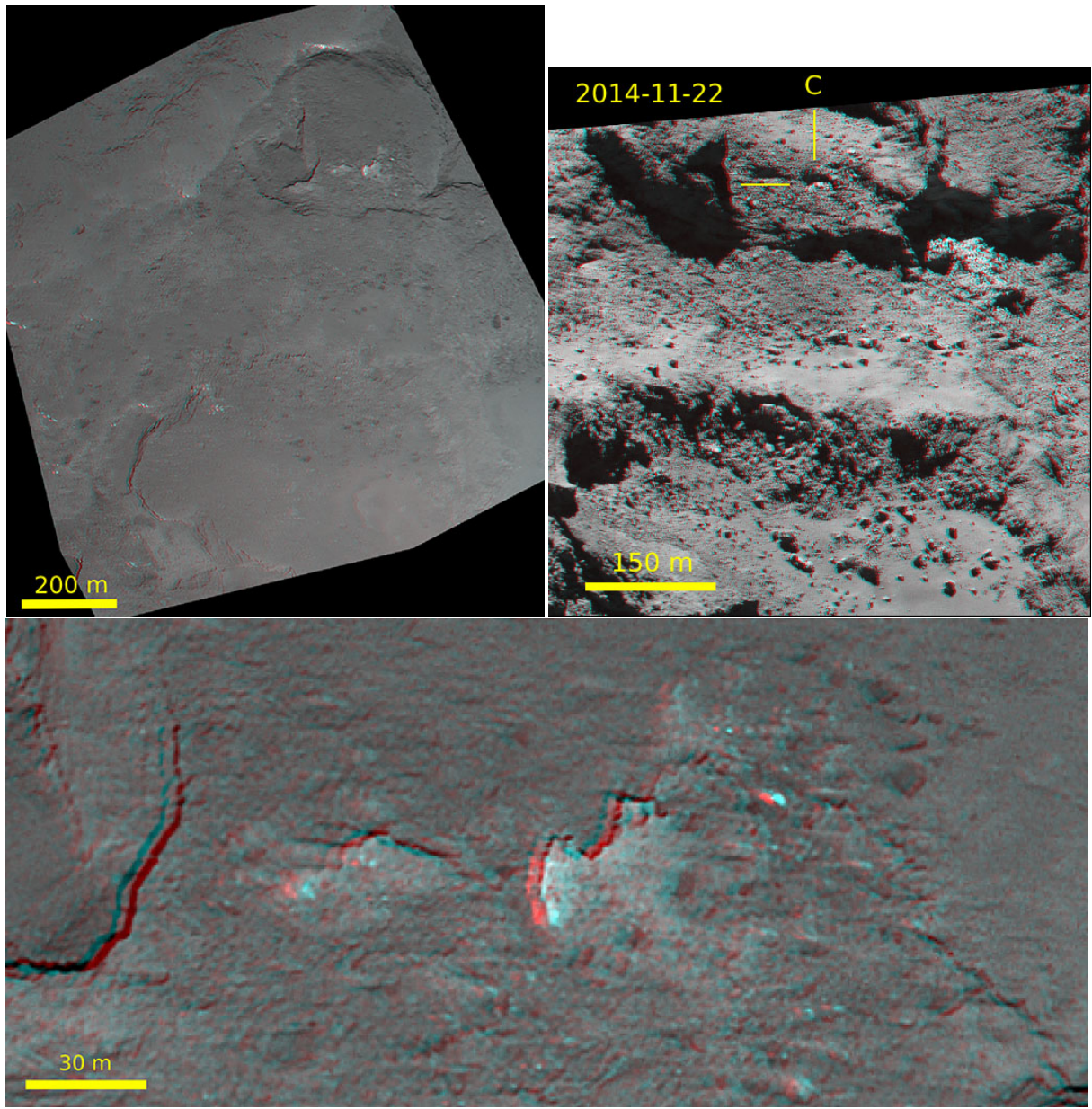

**Figure 5.** Three anaglyphs of cavity $C_C$ at low (upper left panel), medium (upper right panel), and high (lower panel) spatial resolutions.

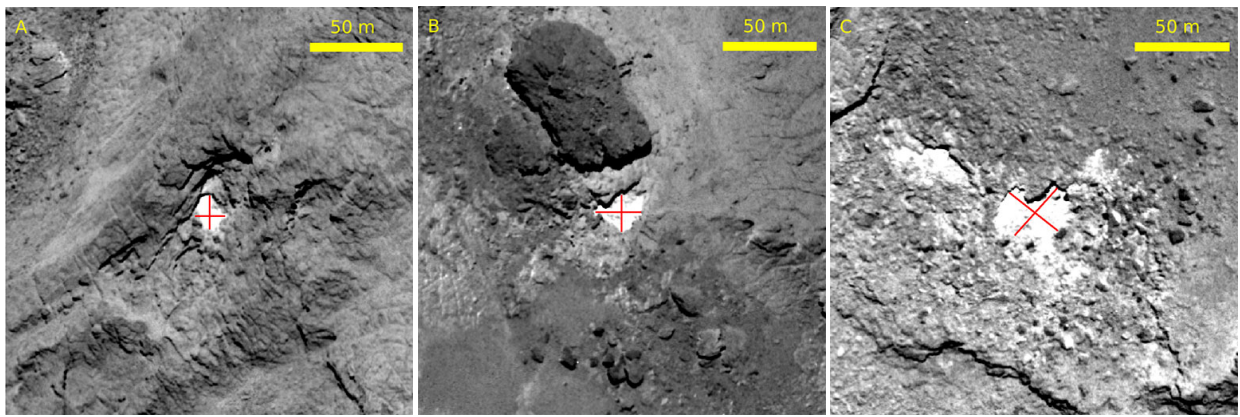

**Figure 6.** Close-up images of the three cavities from the April 2016 sequence used for their two-dimensional characterization. The linear sizes reported in Table 3 correspond to the red segments.

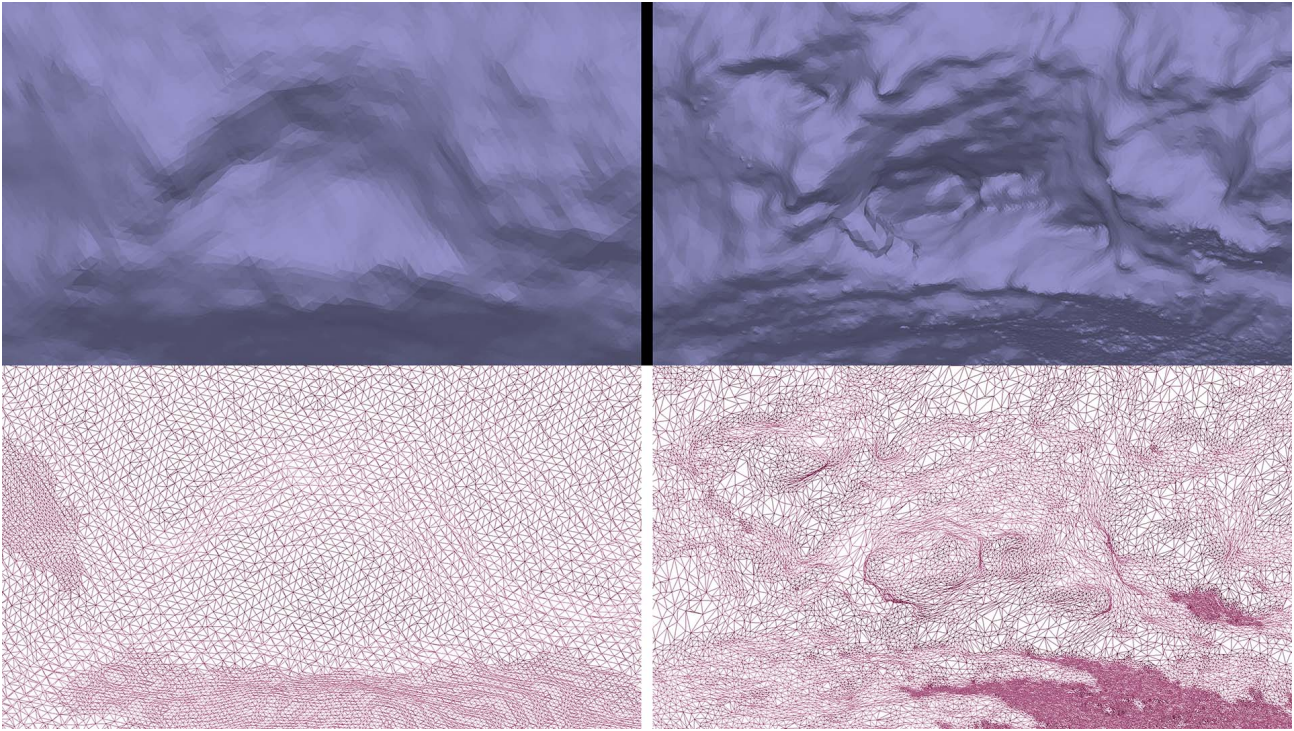

**Figure 7.** Comparison of the SHAP7 model of Preusker et al. (2017) (left column) with our "67P-133M" model in the case of cavity $C$A. Two representations are displayed, solid model (upper row) and wire-frame model (lower row).

Finally, we present a spectacular visit of the cavities and surrounding regions in two movies available on YouTube. Using our high resolution 3D model of the nucleus and the procedure described in Appendix A, we calculated images for a camera flying over the nucleus. The movies are composed of 11 sequences at different distances from the surface. The first one[3] offers a monoscopic vision of the nucleus from a single camera. The second one[4] offers a stereoscopic vision of from a pair of identical cameras. In order to get undistorted images, we applied an axonometric projection with normal 1:1 isometric perspective. For each one, we imposed a parallax of 2.6° appropriate to a comfortable view using a screen of a computer. The stereoscopic base, i.e. the distance between the two cameras, was chosen to ensure a close and comfortable view of stereo-window of the cavity, but remained constant throughout the sequence. This corresponds to the case of human vision since the distance between our eyes (the stereo base) does not vary with the distance to the object.

## 5 PHOTOMETRIC AND SPECTROPHOTOMETRIC PROPERTIES OF THE CAVITIES

The photometric and spectrophotometric analysis started from the spectral radiance $I_{meas}$ in units of $W\,sr^{-1}\,m^{-2}\,nm^{-1}$ given in the first layer of the archive images that we transformed in maps of the radiance factor $I/F_{meas}$, a standard quantity in disk-resolved photometry of solar system airless bodies. It is defined by the following equation:

$$I/F_{meas} = \frac{\pi\, r_h^2\, I_{meas}}{F_\odot(\lambda)},\qquad(1)$$

where $F_\odot(\lambda)$ is the solar spectral irradiance at 1 AU measured at the central wavelength of each filter as given in Table 1, and $r_h$ is the heliocentric distance of the comet in AU.

We selected multi-spectral observational sequences that offer optimal viewings of the cavities for a total of nine images and extracted sub-frames of 400 × 400 pixels centered on the cavities. The log of these nine images is given in Table 4 and it includes the mean and extreme values of the phase angle $\alpha$ as determined over the selected sub-frames for each cavity.

An analysis based on the Hakpe model (Hapke 1993) is beyond the scope of the present investigation and anyway precluded by the limited number of images at the appropriate spatial resolution and by the narrow coverage in phase angle. In view of this last limitation (see last column of Table 4), a simple photometric model restricted to a disk function could a priori be appropriate to correct the radiance factor for the topography and in turn, for the illumination and viewing geometries. As elaborated in Appendix B, we have strong reservations on this procedure and in particular, on the use of the Lommel–Seeliger disk law and briefly summarize the arguments below.

• The disk function which is a function of incident and emission

---

[3] https://youtu.be/ADDyB76qmBI
[4] https://youtu.be/twsfRI52HZw

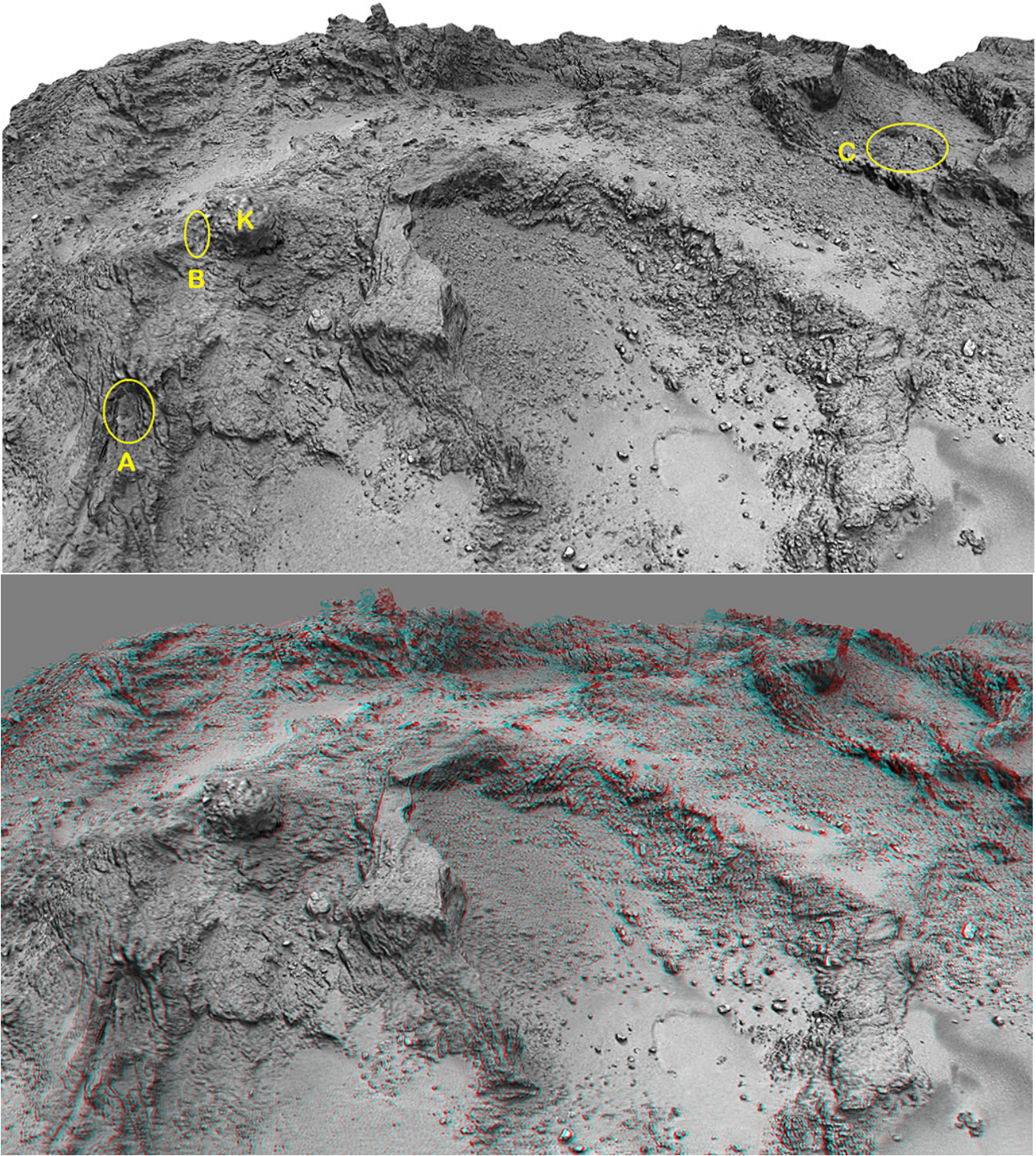

**Figure 8.** Overviews of a region of the nucleus of comet 67P encompassing the three cavities $C_A$, $C_B$, and $C_C$ produced from the 67P-133M shape model. The Kemour mound overlooking cavity $C_B$ is labeled "K". The upper panel presents a two-dimensional view and the lower panel presents an anaglyph of the same region.

angles requires that the pixel corresponds to one or just a few facets and this is far from being the case with the SHAP7 shape model; in addition, this model has intrinsic uncertainties which could propagate to the angles.

- It is highly doubtful that a single law can apply to a large

diversity of terrains including icy patches such as observed on the nucleus of 67P.

- The Lommel–Seeliger disk law is based on radiative transfer which assumes an homogeneous surface. This is certainly not the case of large extents of the nucleus of 67P pervaded by a variety of deposits.

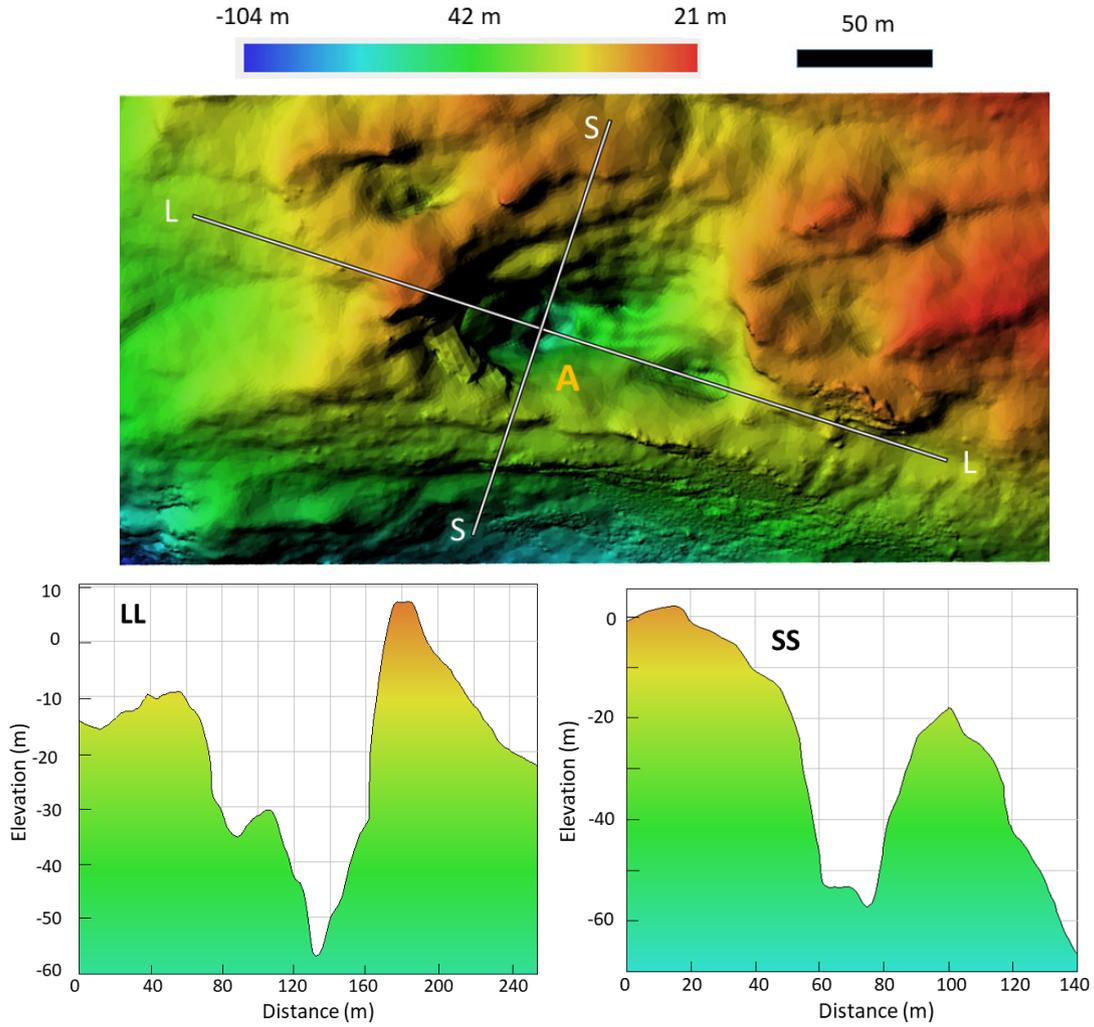

**Figure 9.** Digital elevation model of cavity $C_A$ (upper panel) and profiles across the LL and SS directions (lower panels).

**Table 4.** Log of the images used for the spectrophotometric analysis. $\alpha$, $\alpha_{min}$, and $\alpha_{max}$ are the mean and extreme values of the phase angle for each image.

| Cavity | Date (Apr 2016) | Time | Filter | $\alpha_m$ (deg) | $\alpha_{min} - \alpha_{max}$ (deg) |
|--------|-----------------|------|--------|-----------------|--------------------------------------|
| $C_A$ | 09 | 23:33 | F88 | 4.01 | $2.85 - 5.19$ |
| $C_A$ | 09 | 23:34 | F82 | 3.98 | $2.83 - 5.17$ |
| $C_A$ | 09 | 23:34 | F84 | 3.96 | $2.80 - 5.15$ |
| $C_B$ | 10 | 00:36 | F88 | 5.19 | $4.04 - 6.35$ |
| $C_B$ | 10 | 00:37 | F82 | 5.21 | $4.07 - 6.37$ |
| $C_B$ | 10 | 00:37 | F84 | 5.24 | $4.09 - 6.40$ |
| $C_C$ | 09 | 23:51 | F88 | 1.83 | $0.41 - 3.19$ |
| $C_C$ | 09 | 23:52 | F82 | 1.81 | $0.38 - 3.16$ |
| $C_C$ | 09 | 23:52 | F84 | 1.79 | $0.36 - 3.14$ |

We performed a test in the case of cavity $C_B$ which further revealed additional problems as illustrated in Figure 12. The maps of the incidence and emission angles as extracted from the layers of the archive images show that they are markedly offset from the $I/F$ image and that they present anomalous discontinuities in the contour of the mound. Both discrepancies obviously concur to the introduction of artifacts in the corrected $I/F$ image. On the flip side, we note that the general level at large is barely affected by the correction, the corrected

$I/F$ image being thus very slightly brighter than the original one by typically 3 per cent. In view of these results, we decided against applying the correction. A further justification will appear below as the characterisation of the icy patches prominently relies on the slopes of the spectra which are independent of the geometric correction.

We restrict the presentation of the $I/F$ maps to those obtained with the F88 red filter as they offer the best spatial resolution and as this channel maximizes the contrast of the icy patches with respect to the surrounding terrains (Figure 13). For each patch, the range of $I/F$ was optimized so as to best reveal the photometric variegation across its surface. This is quantified by the frequency distributions of their $I/F$ values as shown in Figure 14: they basically range from 6 to 8% for both $C_A$ and $C_B$, but increase to 6.5 to 8.5% for $C_C$. The peaks of the distributions are reached at nearly the same value of $I/F$ for $C_A$ (6.75%) and $C_B$ (6.83%), but at 7.37% for $C_C$ and this corresponds to a significant ratio of $\approx 1.09$. The $I/F$ uncertainties are discussed in the OSIRIS calibration article by Tubiana et al. (2015): for the visible and near-IR filters of relevance here, they lie in the range 1–1.7 %.

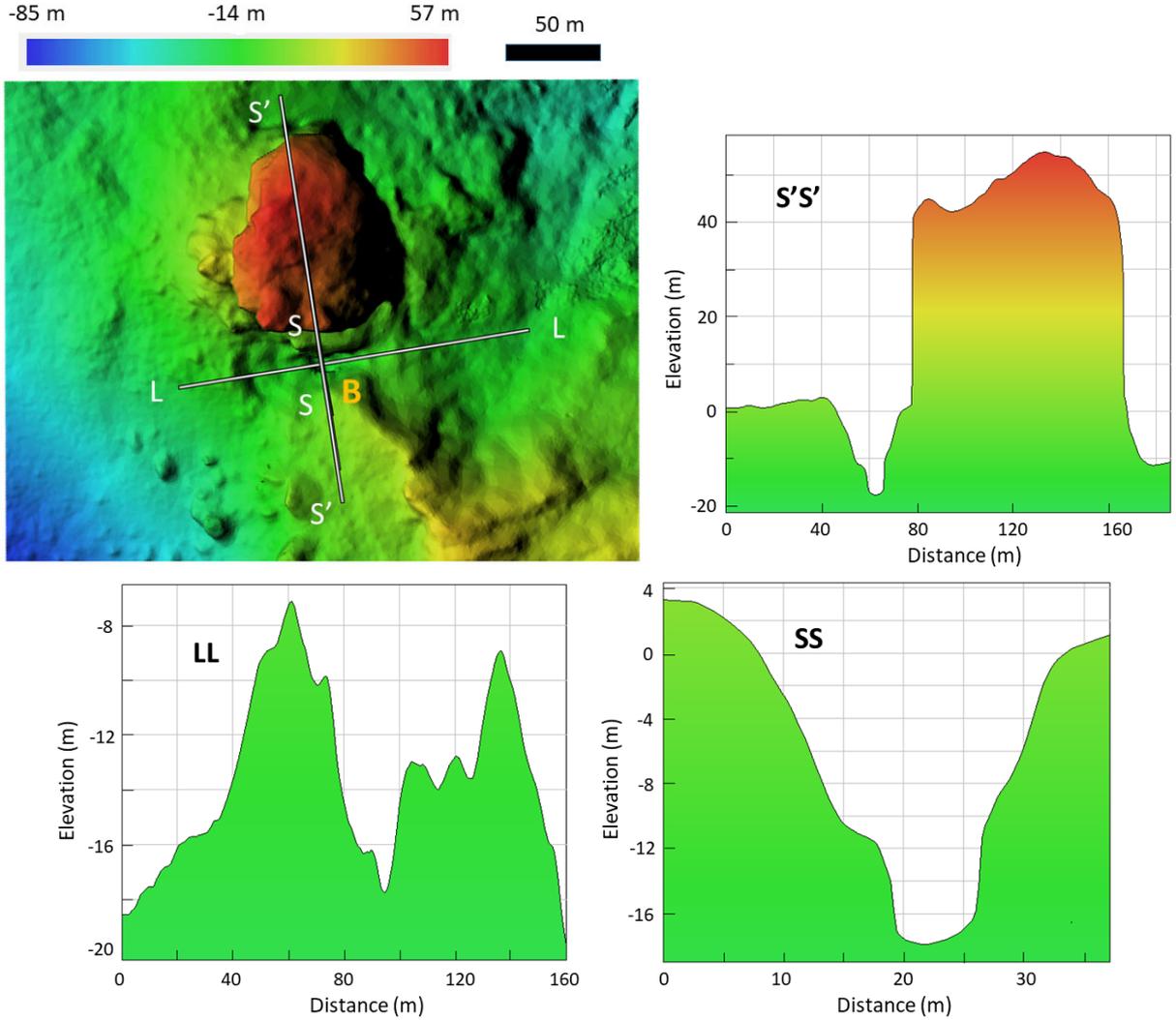

**Figure 10.** Digital elevation model of cavity $C_B$ (upper panel) and profiles across the LL, SS, and S'S' directions (lower panels).

The spectrophotometric analysis started with the accurate co-registration of each set of three images obtained with the three filters. This was performed with the MATLAB "imregcorr" function using the "similarity" option. It estimates the geometric transformation, translation and rotation, which optimally aligns the images. The results are presented by several plots of the relative spectral reflectance which is simply the quantity $I/F$ normalized at a wavelength of 535 nm, here again a standard practice in past articles. The $I/F$ values at this wavelength were calculated by linear interpolations between those at 480.7 nm (F84) and 649.2 nm (F82). In Figure 13, the mean values calculated over the whole area of the icy patches are first compared in with those calculated in four circular apertures surrounding the cavities as depicted by coloured circles, all having a radius of seven pixels. These regions were selected to represent terrains of different morphologies. Following Feller et al. (2019), the uncertainties were estimated by taking the dispersions of values within the respective apertures and they turned out to be < 1 %, therefore smaller than the symbols in Figure 13. The spectra of cavities $C_A$ and $C_B$ clearly stand out by their shallower slopes in comparison with those obtained in the above apertures which themselves do not exhibit much difference; there is an exception in the case of the region surrounding $C_B$ where a slight dispersion of the values appears, prominently at 743.7 nm, but however limited to 2.6%. The spectrum of cavity $C_C$ is slightly steeper than those of the other two cavities, somehow intermediate between those of $C_A$ and $C_B$ on the one hand and those of the surrounding terrains on the other hand.

We refined our spectrophotometric characterization of the icy patches by comparing the mean spectra calculated over their whole area as already presented in Figure 13 with those of two small sub-regions within the patches defined by circular apertures having a radius of three pixels and selected on the basis of their extreme (low and high) brightnesses. Their location and the corresponding spectra are displayed in Figure 15. There is a conspicuous trend of the spectra of the bright spots to be less steep than those of the dark ones and this will be discussed below in terms of the fraction of water ice.

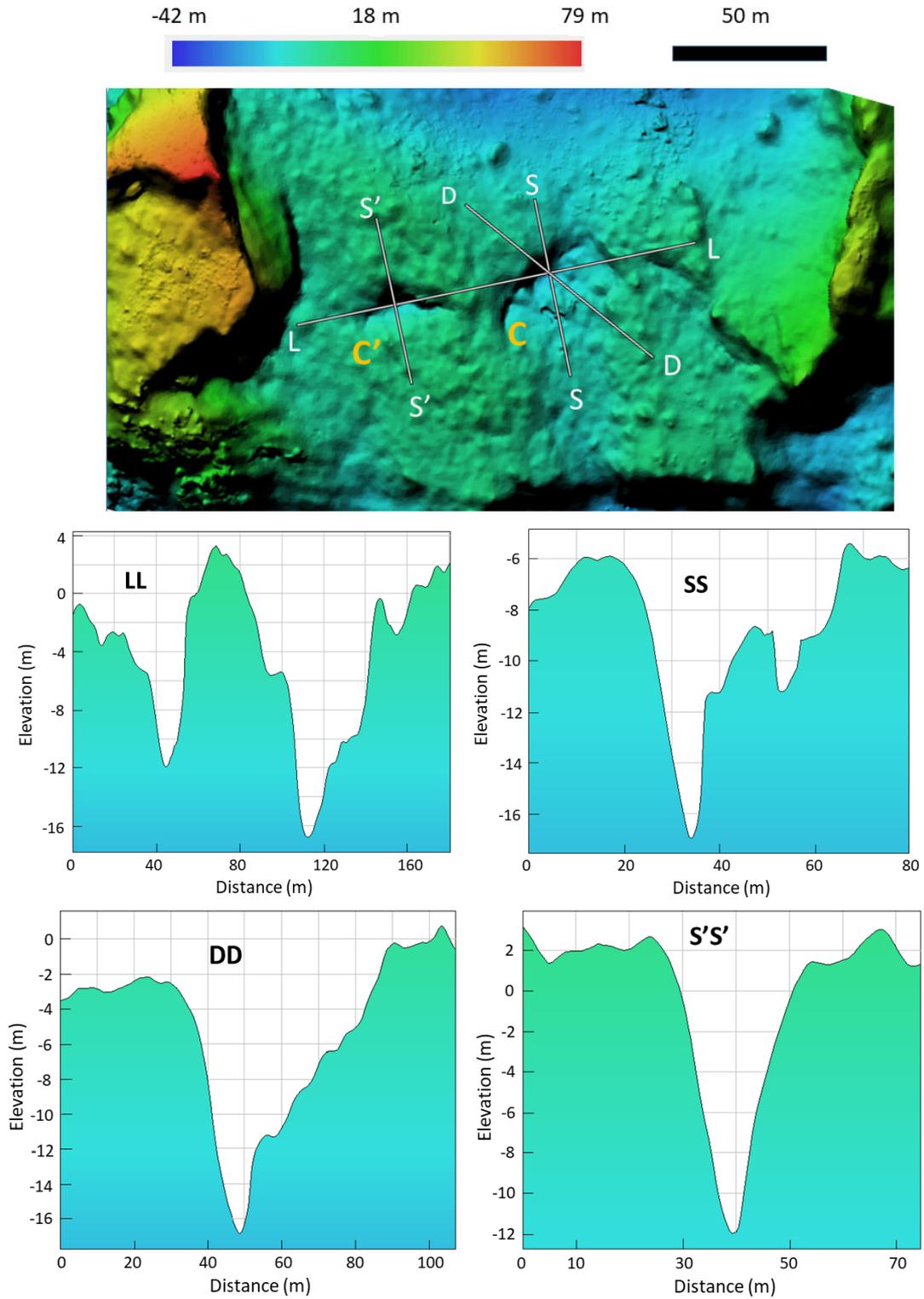

**Figure 11.** Digital elevation model of cavity $C$c (upper panel) and profiles across the LL, SS, S'S', and DD directions (lower panels).

In the case of cometary nuclei and other icy bodies of the solar system, the slopes of the visual-to-infrared (VIS−IR) spectra are commonly interpreted in terms of the relative contributions of water ice and refractory materials (minerals, organic compounds).

According to laboratory measurements and modeling based on the Hapke reflectance model, Ciarniello et al. (2021) showed that the ice abundance is indeed better constrained by spectral indicators such as slopes and band depths than by the VIS absolute reflectance which is not necessarily monotonically linked to ice abundance. Spectral slopes between two wavelengths $\lambda_1$ and $\lambda_2$ as a convenient char-

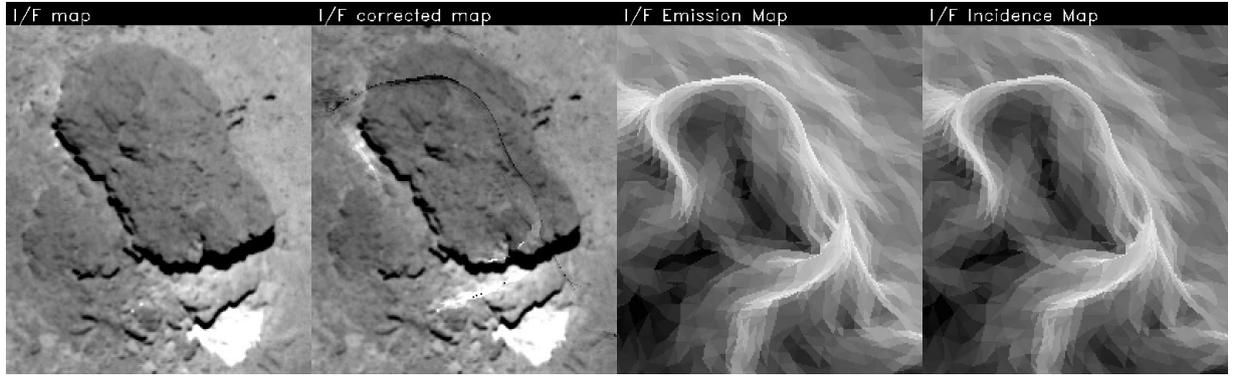

**Figure 12.** Illustration of the test of the application of the photometric correction using the Lommel–Seeliger disk law. From left to right: the $I/F$ original map, the same map corrected by this law, and the maps of the emission and incidence angles.

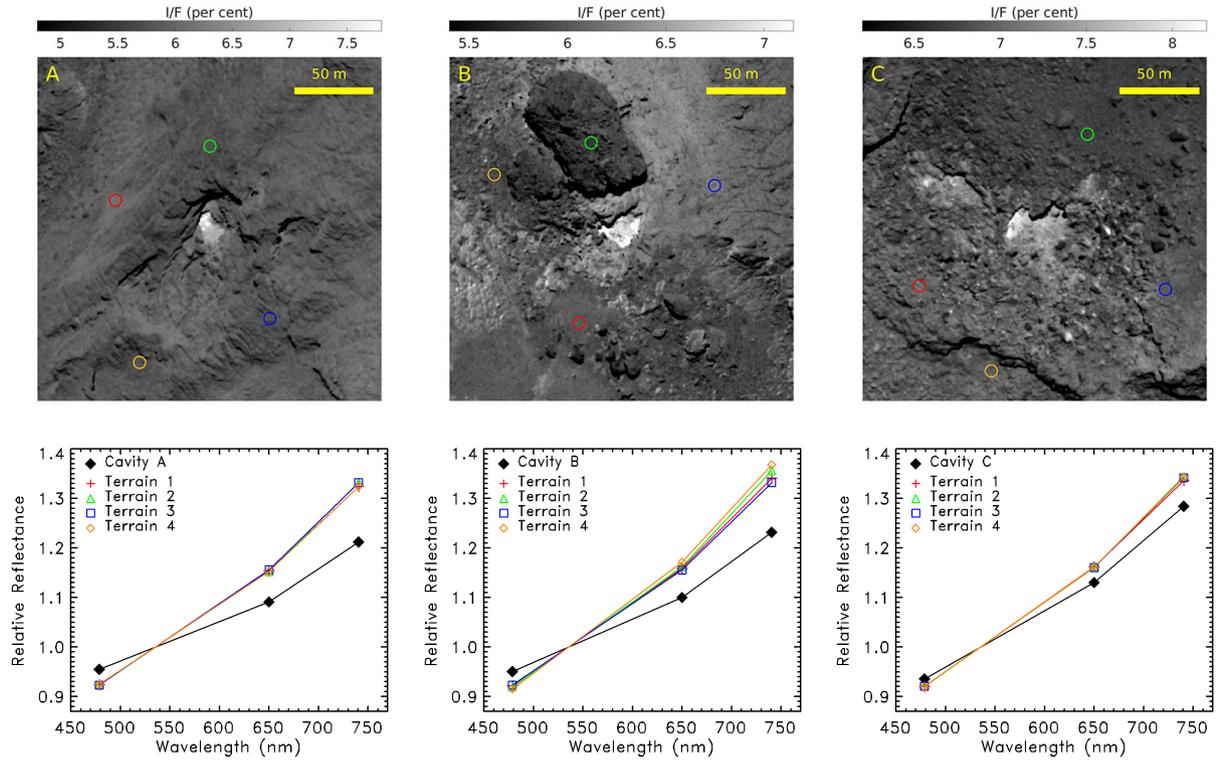

**Figure 13.** Upper row: maps of the radiance factor of the three cavities and surrounding terrains obtained with the F88 red filter. The colored circles correspond to the circular apertures having a radius of 7 pixels where the mean values of the radiance factor were calculated.

Lower row: Spectra of the icy patches integrated over their whole area (black lines) and those of four circular apertures surrounding them (lines **and symbols** of same colors). Observational details are given in Table 4.

acterization of cometary spectra were first introduced by A'Hearn et al. (1984) and became widely used in cometary photometry (Jewitt 1991). Following these authors, the slope $\mathcal{S}$ expressed in units of per cent per 100 nm (abbreviated to % per 100 nm) is given by the following equation:

$$\mathcal{S} = \frac{R(\lambda_2) - R(\lambda_1)}{R_{mean}} \cdot \frac{10^4}{\lambda_2 - \lambda_1} \qquad (2)$$

where $\lambda$ must be expressed in units of nm. $R_{mean}$ is the mean reflectance in the wavelength interval, usually the mid-value of the

extreme reflectances $R(\lambda_1)$ and $R(\lambda_2)$, that is:

$$R_{mean} = \frac{R(\lambda_1) + R(\lambda_2)}{2}. \qquad (3)$$

We draw attention to an alternative definition of $R_{mean}$ incomprehensibly introduced by Fornasier et al. (2015) in the context of OSIRIS where it is replaced by $R(\lambda_1)$. This ends up in the confused situation where some authors used this latter definition (e.g. Feller et al. (2019)) whereas others used the original one (e.g. La Forgia et al. (2015)) thus making comparisons problematic and prone to incorrect conclusions. Here we stick to the original definition as given by Equation 3.

Our results for the spectral slopes calculated in the two wavelength intervals 480.7 – 649.2 and 649.2 – 743.7 nm are presented in Table 5. We concentrate on the second spectral range as it allows comparisons with model spectra to be presented in the next section. A convenient summary if offered by Figure 16 and we see that the mean value of the slope over the overall area of the patches ranges from a low value of 11.6 for $C$A to a high value of 14.0 for $C$C, with an intermediate value of 12.5 %/(100 nm) for $C$B. The twelve values of the surrounding terrains were averaged to produce a "typical" cometary value of 16.0 %/(100 nm), clearly larger than those of the ice patches. The extreme values pretty much follow the same pattern.

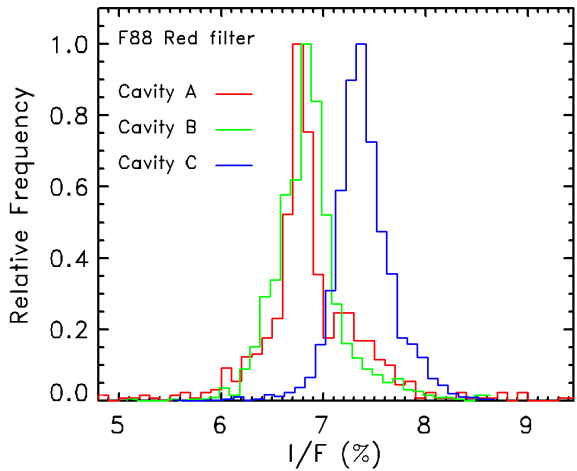

**Figure 14.** Frequency distribution of the radiance factor at 743.7 nm (F88 filter) of the three icy patches.

Figure 17 displays the maps of the spectral slopes in the wavelength interval 649.2 – 743.7 nm of the three ice patches complemented by the corresponding frequency distributions in Figure 18. Cavities $C$A and $C$B may be characterized by similar broad distributions of slopes extending from 0 to 20 %/(100 nm). That of $C$C is conspicuously offset from these last two by approximately +2 %/(100 nm) with further a narrower width extending from 7 to 18 %/(100 nm).

## 6 WATER ICE ABUNDANCE IN THE CAVITIES

The interpretation of these spectral indicators in the case of the nucleus of 67P was prominently developed for the interpretation of the observations obtained by the Visible InfraRed and Thermal Imaging Spectrometer (VIRTIS) aboard the *Rosetta* spacecraft, notably by De Sanctis et al. (2015), Filacchione et al. (2016), Raponi et al. (2016), and Ciarniello et al. (2016). The work of Raponi et al. (2016) is particularly relevant to our analysis as these authors modeled the reflectance of various combinations of crystalline water ice and refractory dark material of the nucleus surface as a function of abundance, grain size, and two mixing conditions: intimate and areal. In the first case, the grains of the two materials are in contact and the light scattering process involves both of them. In the second case, the ice and the dark terrain are in the form of separate patches and the light scattering process is limited inside each patch. For their quantitative analysis, Raponi et al. (2016) implemented the radiative transfer model of Hapke (2012) as described by Ciarniello et al. (2011). Of direct interest to our interpretation is the collection of simulated reflectance spectra, particularly in the wavelength interval 500 – 800 nm. All spectra exhibit a linear part beyond ≈ 650 nm whose slope is therefore well defined and directly comparable to our slopes calculated in the wavelength interval 649.2 – 743.7 nm. The spectral slopes of the models were estimated from figure 8 of Raponi et al. (2016) and are summarized in Table 6. As pointed out by these authors, the spectral slopes of the mixtures of water ice and dark material in the visible domain are insensitive to the size of the grains so that they only put constrains on the abundances. In order to derive them from our slopes, we fitted simple functions to the data points of Table 6, a linear one in the case of intimate mixing and a quadratic one in that of areal mixing (Figure 19). This allowed transforming the slope maps to ice abundance maps as displayed in Figure 17.

The choice between intimate and areal mixtures is not clear cut and, in the latter alternative, is a matter of spatial scale of the individual patches of the two materials. The histograms of the slope over the cavities indicate continuous distributions at the pixel scale (typically 50 cm) rather than bi-modal distributions (Figure 18), a trend against areal mixing. However, individual patches at lower scales but still much larger than visible wavelengths thus isolating the light scattering process cannot be ruled out. Whatever the case, we present in Figure 17 the maps of the ice abundance in the three patches for the two types of mixture as well as their frequency distributions in Figure 20. They do not exceed a few per cent and are systematically smaller in the case of areal mixing than in the case of intimate mixing. Considering altogether $C$A and $C$B, they range between 0 and 2.5 % in the former case and between 0 and 10 % in the latter case. The abundances are substantially lower for $C$C, the corresponding ranges being 0 to 1 % and 0 to 6 %, respectively.

Whereas the determination of the spectral slopes is quite robust, the above values of the ice abundance are model-dependent further limited by i) the small number of models and ii) our interpolations between the model data of Raponi et al. (2016). Nevertheless the low abundances of water ice in the bright patches of 67P, typically a few per cent, are fully consistent with the results derived from the VIRTIS observations which, in addition to the visible slopes, are further constrained by the infrared spectral indicators, prominently the ice absorption bands.

Finally, we note that one of the three large patches "BAPs" (bright albedo patches), namely 'BAP 1' extensively studied by Raponi et al. (2016) based on VIRTIS observations during the last four months of 2014 is located close to our cavity $C$C. Their latitudes are indeed very close, but the longitude difference of nearly 8° rules out that they were the same feature.

## 7 LIFETIME OF THE ICY CAVITIES AND CONNECTION TO COMETARY ACTIVITY

The lifetime of the three icy cavities was constrained by investigating images and anaglyphs before and after their discovery during the 2016 April 9-10 observational sequence, keeping in mind that its determinations may well be hampered and biased by the observing conditions. This is particularly the case of $C$A which turned out to be in the shadow of the scarp in all observations except precisely during the April sequence. The same situation pretty much prevails for $C$B as a consequence of the shadow cast by the mound. However, a bright spot at the right location is present in images of 2014 September 16 and later on an image of 2016 January 30, both detections likely resulting from a small fraction of the cavity being unobstructed by the mound or its shadow. After the sequence of April 2016, $C$B is well visible on images of 2016 July 2. Altogether, this suggests a possible lifetime of almost two years, well in line with results from other investigations, notably Oklay et al. (2017). In $C$C a few bright spots are already visible on images of 2014 September. A spectacular anaglyph of 2016 June 15 shown in Figure 21 reveals that $C_C$ was then almost covered with debris leaving only a small light-grey patch. This allows setting an upper limit to the lifetime of $C_C$ at 1.8 yr.

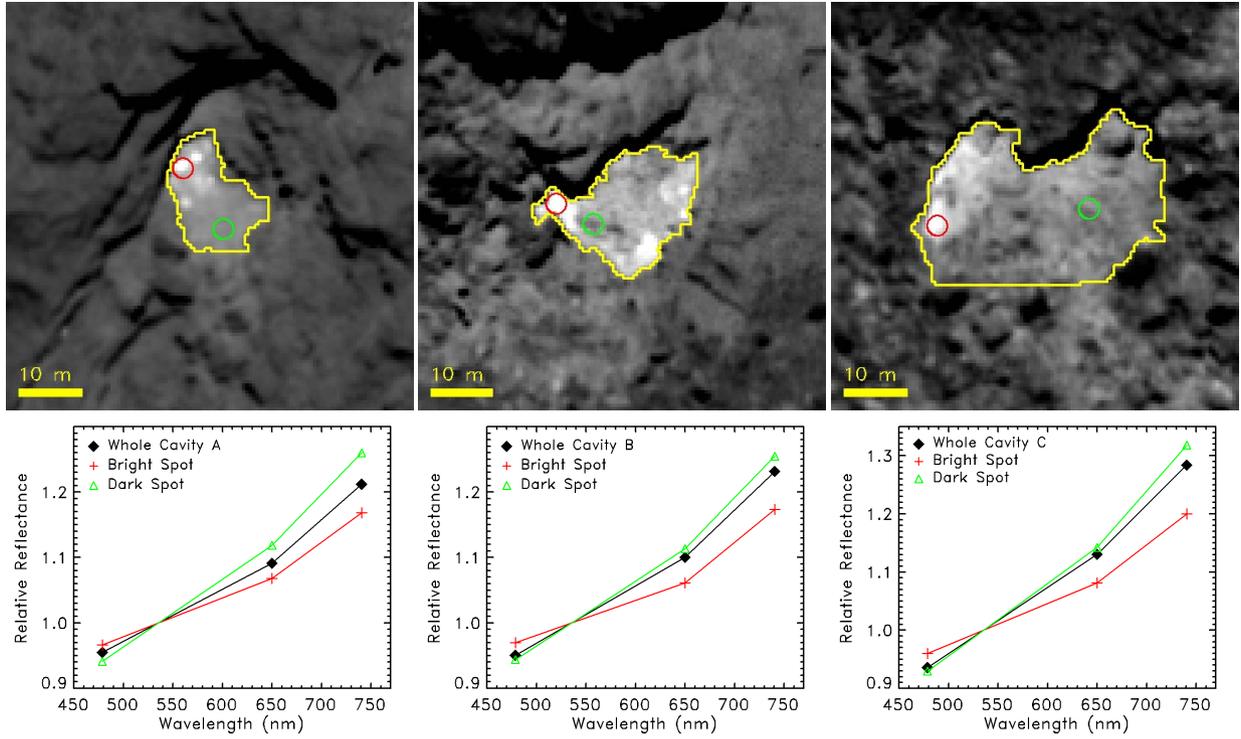

**Figure 15.** Upper row: maps of the radiance factor of the three cavities showing the location of the circular apertures having a radius of 3 pixels where the mean values were calculated. The yellow contours delimit the areas used for the maps displayed in Figure 17. Lower row: Spectra of the icy patches over their whole area (black lines) and those calculated in the above apertures (lines and symbols of same colors).

**Table 5.** Spectral slopes in the two wavelength intervals 480.7 – 649.2 nm [F84-F82] and 649.2 – 743.7 nm [F82-F88] expressed in units of ‰/(100 nm). For each of the three ice patches, columns #2 to #4 give the mean value of the slope over its overall area and the two mean values over the small sub-regions (circular apertures) selected for their extreme brighnesses, low (subscript "l") and high (subscript "h"). The last four columns labeled "T1" to "T4" list the slopes in the four circular apertures surrounding the cavities as defined in Figure 13

| Cavity & terrain | $C_A$ | $C_{A,l}$ | $C_{A,h}$ | T1 | T2 | T3 | T4 |
|---|---|---|---|---|---|---|---|
| Slope [F84-F82] | 7.8 | 10.1 | 5.9 | 12.8 | 12.9 | 13.1 | 12.8 |
| Slope [F82-F88] | 11.6 | 12.9 | 10.1 | 15.3 | 15.9 | 15.6 | 15.3 |
| | $C_B$ | $C_{B,l}$ | $C_{B,h}$ | T1 | T2 | T3 | T4 |
| Slope [F84-F82] | 8.5 | 9.5 | 5.1 | 13.3 | 13.6 | 13.1 | 14.4 |
| Slope [F82-F88] | 12.5 | 13.7 | 11.4 | 16.1 | 17.1 | 15.7 | 17.0 |
| | $C_C$ | $C_{C,l}$ | $C_{C,h}$ | T1 | T2 | T3 | T4 |
| Slope [F84-F82] | 11.0 | 12.0 | 7.0 | 13.5 | 13.4 | 13.5 | 13.5 |
| Slope [F82-F88] | 14.0 | 15.8 | 11.4 | 15.6 | 16.4 | 16.0 | 16.2 |

We searched whether our three cavities were mentioned in past investigations of bright spots and sources of jets. The map of active sources during April 2015 constructed by Vincent et al. (2016) presented in their fig. A.5 reveals that the location of their source #24 is in excellent agreement with our cavity $C_B$. This implies not only that it was present at that time in agreement with our conclusion on its lifetime, but already active.

Our three cavities were identified by Feller et al. (2019) which is not a surprise since they analyzed the same observational sequence. They appear as red spots in their maps displayed in their fig. 2, 4, 5, and 9, but only one of them, namely $C_A$, received in-depth consideration. It is listed as feature BF-02 in their table A.2 which gives its position in a different coordinate system having its origin at the Cheops boulder. Translating to the official system of Preusker et al. (2017), we found an excellent agreement with our determination. Curiously, the same feature received a different code, BF-A002, in their fig. 8 which displays their spectral measurements.

**Table 6.** Spectral slopes in the wavelength interval $650-800\,$nm expressed in units of $\%/(100\,$nm$)$ of various models constructed by [Raponi et al. (2016)](#) combining water ice and refractory dark material. The parameters ice abundance and size of the ice grains are allowed to vary one at a time. In the case of intimate mixing, when the abundance of ice varies, the grain diameter is fixed at 50 μm; conversely, when the diameter varies, the abundance is fixed at 2 %. In the case of intimate mixing, when the abundance of ice varies, the grain diameter is fixed at 2000 μm; conversely, when the diameter varies, the abundance is fixed at 0.5 %.

| Mixing | Intimate | | | Areal | | |
|---|---|---|---|---|---|---|
| Ice abundance (%) | 1.0 | 2.0 | 3.0 | 0.0 | 0.5 | 1.0 |
| Spectral slope | 12.0 | 11.0 | 10.0 | 14.8 | 11.0 | 8.4 |
| Grain diameter ( μm) | 20 | 50 | 200 | 1000 | 2000 | 3000 |
| Spectral slope | 11.1 | 11.1 | 11.1 | 11.7 | 11.7 | 11.7 |

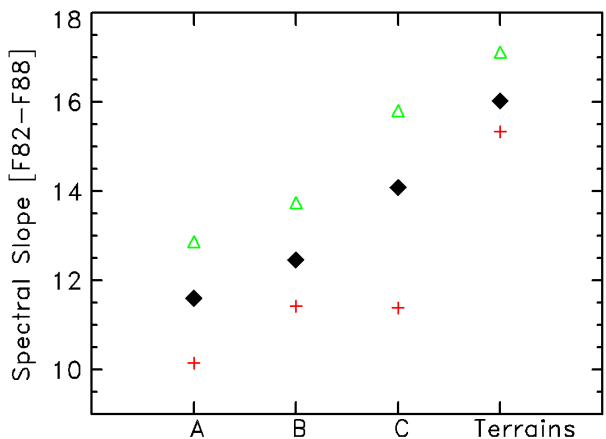

**Figure 16.** Synthesis of the spectral slopes in the wavelength interval 649.2 – 743.7 nm expressed in units of % per 100 nm. For each of the three ice patches, we give the mean value of the slope over its overall area (black diamonds) and the two mean values over the small sub-regions (circular apertures) selected for their extreme brightnesses, low (green triangles) and high (red crosses). For the twelve determinations of the surrounding terrains considered altogether, we indicate the mean (black diamond), maximum (green triangle), and minimum (red cross) values of the slope.

We checked the recent extensive catalogue of [Fornasier et al. (2023)](#) which identifies more than 600 volatile exposures on the nucleus. Out of our three cavities, only one, $C$C, is possibly listed in their table 4 as its coordinates are within 1° of their exposure number 54 (Lat = 14.04°, Lon = 111.08°). However, their area of 5.5 m² is way off our value of 804 m²; however, it may be restricted to the brightest spot of the cavity. There are two exposures located in the Khepry region (two overlapping red discs on their fig. 1, but absent in the list) however a few degrees away from $C$B. Finally, $C$A is totally missing whereas it was inventoried by [Feller et al. (2019)](#).

[Lai et al. (2019)](#) investigated the spatial and temporal variations in source regions of the dust jets as significantly influenced by seasonal effects. Their fig. 4 displays two maps of the jet source regions from pre- and post-perihelion observations. The locations are colour coded, the colour bars allowing a temporal resolution of typically 20 days. In spite of the low spatial resolution of the maps, we could identify our three cavities on the pre-perihelion map: $C$A in March-April 2015 (green disk), $C$B in June-July-August 2015 (three overlapping disks of different colours), and $C$C after 2015 August 8 (red disk). On the post-perihelion map, $C$A is present in October 2015 (cyan disk) and $C$C in November-December 2015 (green disk) whereas $C$B is absent.

[Fornasier et al. (2019)](#) focused on the identification of the source regions of more than 200 jets. A source does appear at the location of $C$B on their map (their fig. 1) as a tiny yellow triangle and is associated with jet #133 in their table A.1. whose coordinates agree with ours within a fraction of a degree. Likewise, another source coincides with $C$C (large yellow triangle in the Ash region) and is associated with jet #42 whose coordinates are in perfect agreement with ours.

We present further evidence of a jet most likely associated with $C$B on 2015 July 18 thanks to a sequence of 16 anaglyphs constructed from 25 OSIRIS-NAC images taken between 21:00:05 and 23:00:05 at a cadence of 5 minutes. Among the forest of permanent, diffuse jets that pervaded the southern hemisphere of the nucleus in July 2015, a transitory bright one appeared at 22:01:23. Figure [22](#) presents three anaglyphs: pre- and post-event views bracket one view of the jet at its culmination. It appears very close to the Kemour mound, but the viewing conditions were such that cavities $C$A and $C$B were nearly aligned making both of them potential candidates as its source. However, careful inspection of the geometry of the jet and the mound indicated that the latter would partly block the visibility of the jet if it originated from $C$A and this is not the case. Therefore, $C$B was favoured as the source of the jet and this will be confirmed on the basis of the illumination of its bottom in the next section.

Likewise the visit of the cavities, we present two movies available on YouTube of the full sequence offering monoscopic[5] and stereoscopic[6] visualizations of the temporal evolution of the jet. However, since the anaglyphs combine two successive images separated by 5 minutes, we went back to the original images to determine a precise time-line of the development of the jet. It started between 21:55:06 (no trace) and 22:00:06 when it was already quite bright and collimated. It then became more diffuse and progressively declined until 22:15:06 when a very faint residual may be perceived. There is no trace left on the next image at 22:20:06. This sets the lifetime of the jet of approximately 20 minutes.

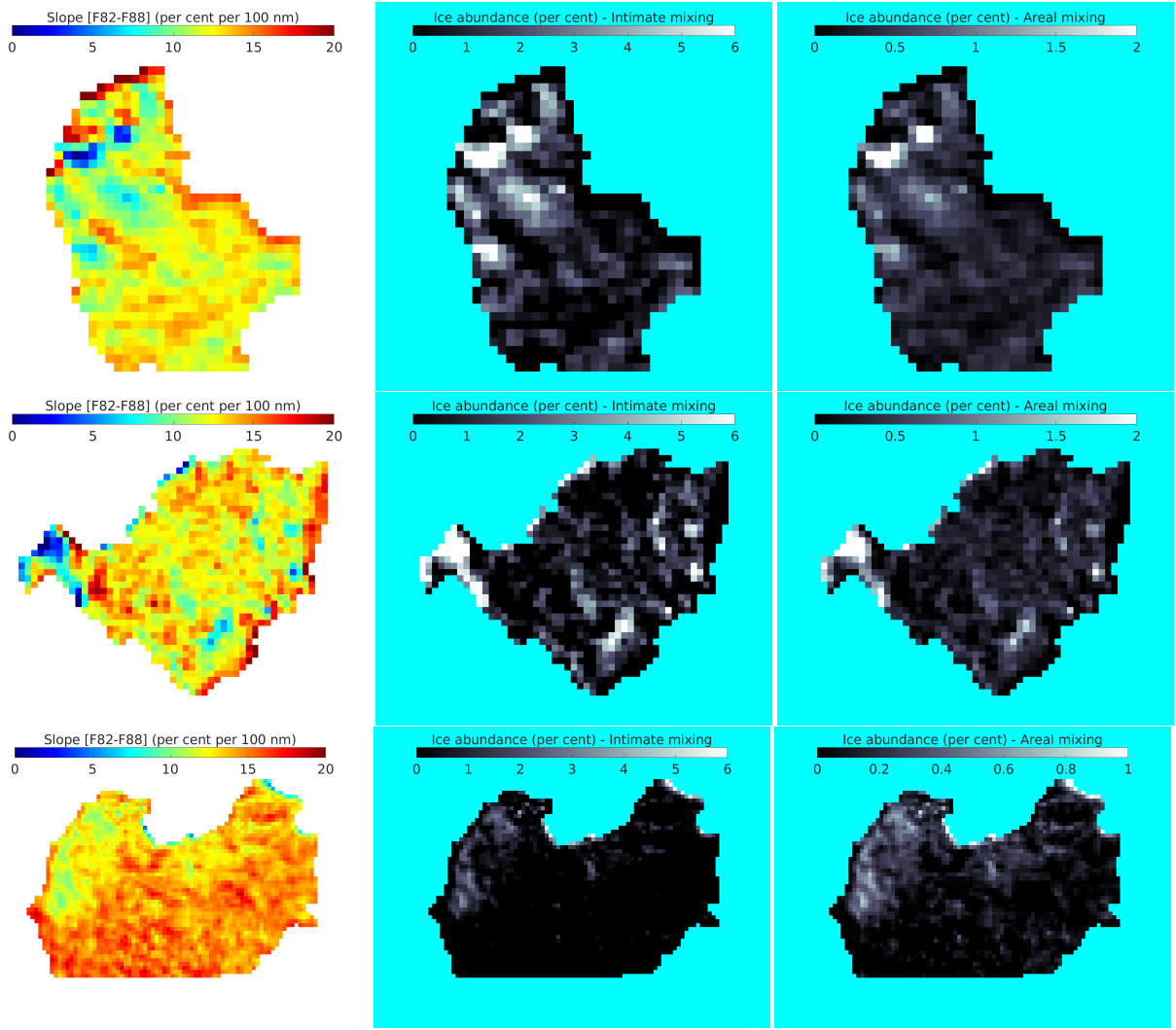

**Figure 17.** Maps of the spectral slopes in the wavelength interval 649.2 – 743.7 nm (left column), maps of the derived ice abundance in the case of intimate mixing (central column), and maps of the derived ice abundance in the case of areal mixing (right column) of the three ice patches: $C_A$ (upper row), $C_B$ (middle row), and $C_C$ (lower row).

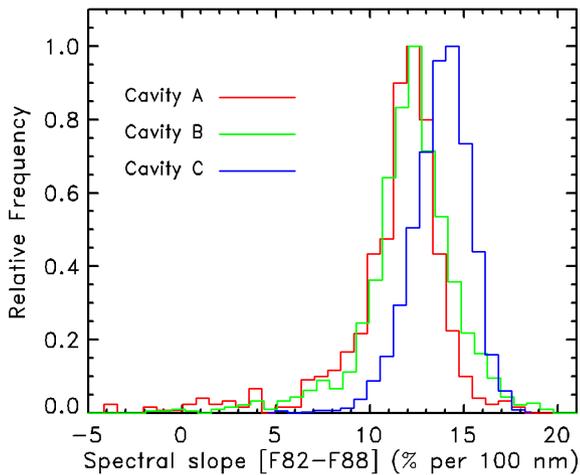

**Figure 18.** Frequency distribution of the spectral slopes in the wavelength interval 649.2 – 743.7 nm [F82-F88] of the three ice patches.

## 8 THERMAL CONSTRAINTS ON THE CAVITIES

In order to investigate and quantify the thermal constraints on the cavities, we implemented a thermal model developed for the nucleus of 67P and described in (Attree et al. 2019). We used a simplified version of this model, which only takes into account solar insolation, surface thermal emission, and projected shadows. The accumulated energy over a complete orbital revolution of the comet received by each facet of the DEMs of the three cavities and surrounding terrains was calculated 36 times per rotation (i.e. every 1240 s) to track the diurnal short term variations and at 35 different times over the orbit separated by variable time interval (from 100 days at aphelion to 15 days at perihelion) to track the seasonal long term variations. At each time step, the distance to the Sun, the orientation of each facet relative to the Sun, and the projected shadows were computed using the OASIS software (Jorda et al. 2010). Solar insolation being independent of the surface composition, the sublimation of water ice was neglected in this model.

The resulting maps are displayed in Figure 23. Their spatial extents and scales are similar to those of the digital elevation models of the three cavities shown in the upper panels of Figure 9, 10, and 11 to facilitate the comparison. As expected, the bottom of the cavities received significantly smaller quantities of energy (typically $4 \times 10^9 \, J \, m^{-2}$) than the well-exposed surrounding terrains (typically 7 to $8 \times 10^9 \, J \, m^{-2}$). Some limited parts of the bottom of $C$A received even lower levels, approximately $2 \times 10^9 \, J \, m^{-2}$. In contrast, $C$C was more exposed to sunlight likely explaining the rapid disappearance of the icy patches underlined in Section 7.

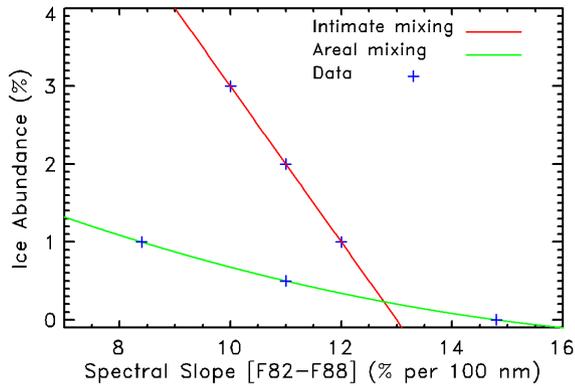

**Figure 19.** Ice abundance versus spectral slope. The data points come from the models of [Raponi et al. (2016)](#). Functional representations are shown by solid lines for the two cases of intimate and areal mixings.

We performed a specific study in the case of $C$A and $C$B to further ascertain which of the two was the source of the jet observed on 2015 July 18 and which lasted approximately 20 minutes. We selected a few facets at each bottom and monitored the respective averaged insolations during the rotational period of the nucleus with a time resolution of 10 minutes. A few facets of the top of the Kemour mound were included as a "witness plate" whose insolation is unobstructed and only modulated by the rotation of the nucleus. The resulting temporal profiles of the surface insolation are displayed in Figure 24 relative to the time of the quasi maximum of the jet, that is 22:00 UT. The extremely short insolation of $C$A as well as its large temporal offset from the occurrence of the jet clearly rule this cavity as a potential source. In contrast, the bottom of $C$B is copiously illuminated, almost as the top of Kemour, but during a shorter time interval due to partial obstruction of the insolation by the walls of the cavity. The thermal lag of the peak of the $C$B profile amounts to only 23 minutes with respect to the quasi maximum of the jet ensuring that it is indeed its source.

## 9 DISCUSSION

Whereas the presence of icy cavities was suspected and mentioned in several past articles, none were characterized as such since this is quasi impossible on the basis of individual images. In addition, the spatial resolution of the best shape model so far (SHAP7) is insufficient to investigate these features. Stereophotography has the outstanding merit of offering a three-dimensional visualization at the highest spatial scale allowed by the images further enhanced by the process of merging the two images performed by the brain. Clearly, the three cavities studied in this article would not have drawn attention beyond the past brief mentions in the absence of anaglyphs which revealed their potential as subsurface access points. On the one hand, stereophotography does offer the best unbiased views of these cavities allowing their physical description, but are unable to provide quantitative information. Our new high resolution shape model supplements this missing information and, of utmost importance, it provides an estimate of the depth of the cavities. It must however be realized that the quality of the information, and particularly of the profiles, very much depends upon the number of suitable images and upon their spatial resolution. With our model, we reached the ultimate step allowed by the full data set of OSIRIS post-perihelion images, but the morphology of the cavities may be more complex than that indicated by the relatively smooth profiles and their depth may even be larger. Here again, anaglyphs come to the rescue, but only in the favourable case of $C$A where the rugged nature of its wall may be perceived as stacks of individual boulders (Figure 3).

The size, depth-to-diameter ratio, and shape of the cavities make them truly distinct from the common pits such as studied by [Vincent et al. (2015)](#) and make them serious candidates as subsurface access points. Their bottom has photometric and spectrophotometric properties that make them distinctly different from the surrounding terrains: larger radiance factors and even more important, shallower spectral slopes. This is indeed an important result as the determination of slopes is very robust, being independent of any photometric corrections. Their interpretation in terms of the presence of water ice could in principle be disputed, but two evidences argue in favour of it. First, the presence of water ice as explaining bright patches is supported by infrared observations, particularly the detection of characteristic absorption bands. Second, the likely association of a jet with $C$B on 2015 July 18 just one month before the perihelion passage of the comet can only be explained by the exposition of water ice to solar illumination as other more volatile ices would have disappeared much earlier. As a bonus, the geometry of the narrow cavity naturally explains the brevity of the jet and its collimation. The derivation of the abundance of water ice from the spectral slopes is undoubtedly model-dependent, but the low values we obtained – typically a few per cent – are fully supported by several evidences coming from infrared observations (e.g. [Raponi et al. (2016)](#)).

The lifetime of the icy bottom of the cavities imposes serious constraints on the origin of the water ice and on the mechanism(s) that formed the cavities themselves. Unfortunately, their determinations are very much biased by the availability of suitable observations, optimally the continuous long-term monitoring of a given region, but this is seldom the case. Recent analysis established a dichotomy between bright spots with lifetimes in the range of minutes to hours associated with temporary frost deposits and those with much longer lifetimes extending over several days to months associated with exposures of original water ice enriched blocks ([Ciarniello et al. (2022)](#); [Fornasier et al. (2023)](#)). Clearly, our analysis of icy cavities shows that a third, alternative origin of the bright spots is possible. We were able to gather evidences of their presence during approximately two years for $C$B and $C$C. This is surprising long, but well in line with other investigations (e.g. [Oklay et al. (2017)](#)). Recurrent frost condensation followed by its sublimation over such a time lapse does not appear realistic. Exposures of water ice enriched blocks as proposed by [Ciarniello et al. (2022)](#) take place at the surface of the nucleus, not in cavities. We are thus strongly inclined in favour of the presence of a layer or of pockets of pristine subsurface mixtures of water ice and refractory grains at the bottom of the cavities, here again supporting their classification as subsurface access points.

Unlike the pits studied by [Vincent et al. (2015)](#) best explained by sinkhole collapse, such a process appears difficult to advocate for the cavities. They are probably native voids in a low density nucleus with inhomogeneities in the first few hundred metres below its current surface as concluded by these authors. This does not contradict the results from gravity field measurements of [Pätzold et al. (2016)](#) who concluded that the interior of the nucleus is homogeneous *on a global scale* without large voids.

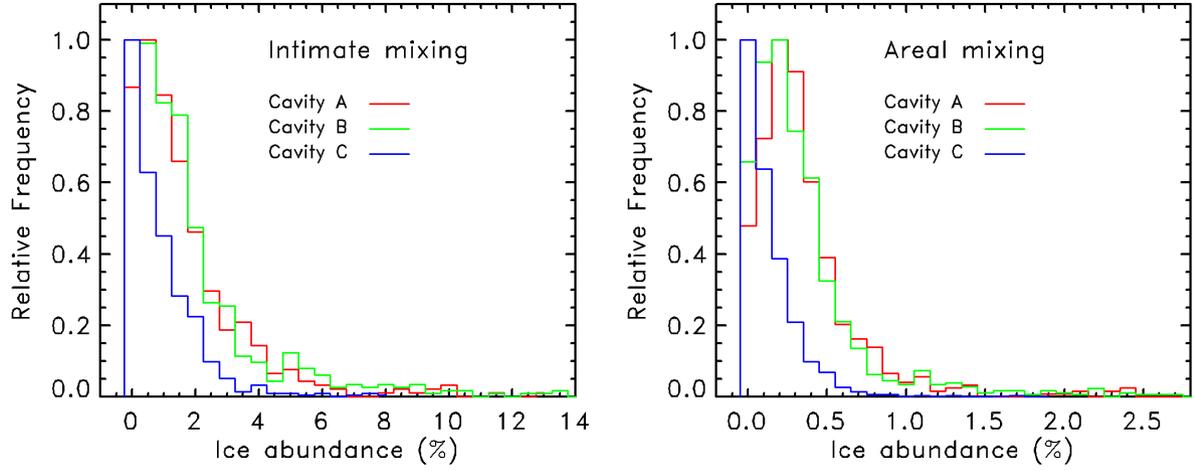

**Figure 20.** Frequency distribution of the ice abundance in the three icy patches for the two cases of mixing: intimate (left panel) and areal (right panel).

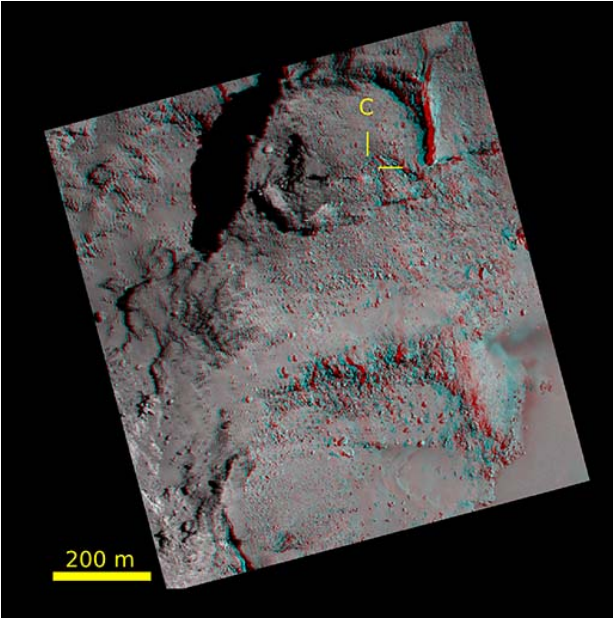

**Figure 21.** [Anaglyph](#) showing the state of cavity $C_C$ on 2016 June 15.

Whereas $C_A$ and $C_B$ are characterized by similar properties, $C_C$ stands out as somehow intermediate between these two cases and the surrounding terrains and as such, reveals the variety of small scale features on the nucleus. Its location in a depression whose floor is undergoing severe degradation (Figure 21) makes interpretation of its formation, as well as that of $C_C'$ even more difficult. It is possible that ice-rich pockets were present underneath the floor before it brook apart, the cavities then resulting from the disappearance of the ice-dust mixture under solar insolation. In that respect, we recall that $C_C$ was associated with jet #42 in the list of [Fornasier et al. (2019)](#) in Section 7.

## 10 CONCLUSION

Our research fits well with recent efforts that recognize the value of stereophotography as a tool for the visualization and the characterization of the surface of solar system bodies at spatial scales which are usually not reached by digital terrain models. It further is part of the developing interest in subsurface access points and their use as a mean of probing the interior of these bodies. We showed for the first time that a cometary nucleus, namely that of comet 67P/Churyumov-Gerasimenko, has cavities that undoubtedly satisfy the criteria of SAPs. In parallel, we produced a new shape model of the nucleus whose spatial resolution surpasses those published so far. Our main findings are summarized below.

(i) We successfully detected and fully characterized three icy cavities on the big lobe of the nucleus of comet 67P/Churyumov-Gerasimenko in regions exposed to sunlight at perihelion time thanks to a set of anaglyphs.

(ii) They visually appear as bright patches of typically 15 to 30 m across whose large reflectances and spectral slopes in the visible substantiate the presence of sub-surface water-ice.

(iii) Our new high-resolution photogrammetric shape model of the nucleus allowed us to construct the profiles of these cavities whose depth ranges from 20 to 47 m.

(iv) Spectral slopes were interpreted with models combining water ice and refractory dark material and considering two cases of mixing, intimate and areal. The abundances of ice do not exceed a few per cent and are systematically smaller in the case of areal mixing than in the case of intimate mixing. Considering altogether cavities $C_A$ and $C_B$, they range between 0 and 2.5 % in the former case and between 0 and 10 % in the latter case. The abundances are substantially lower in cavity $C_C$, the corresponding ranges being 0 to 1 % and 0 to 6 %, respectively.

(v) The determination of the lifetime of the icy cavities is strongly biased by the availability of appropriate and favourable observations, but we found evidences of values of up to two years. Furthermore, these cavities are connected to jets well documented in past studies.

(vi) A thermal model allowed us to track the insolation of the cavities and their surroundings over a large part of the orbit of the comet. As a direct application, we showed that the transitory bright jet that appeared on 2015 July 18 and lasted 20 minutes unambiguously resulted from the brief solar illumination of the icy bottom of the $C_B$ cavity. It is likely the first time that such a direct link is established.

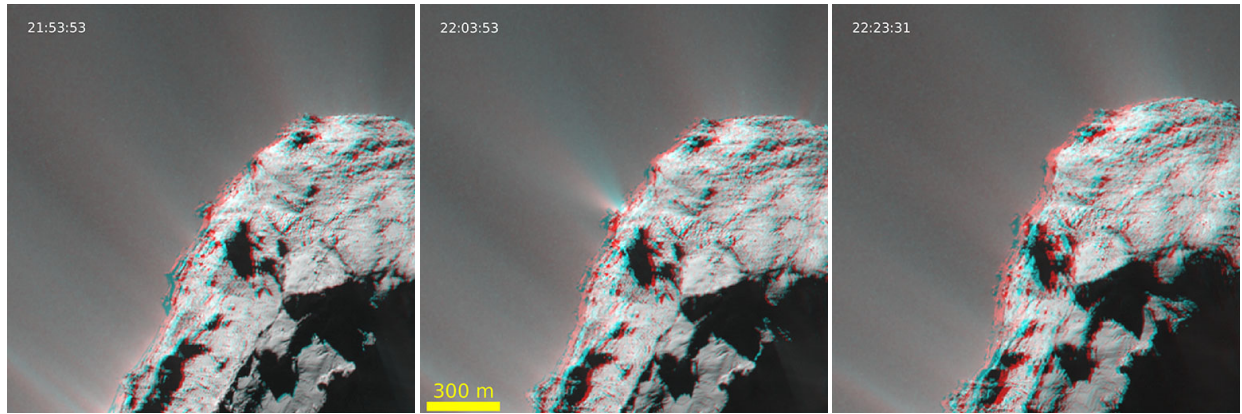

**Figure 22.** Three anaglyphs extracted from the movie available online as supplementary material and illustrating the transitory jet associated with cavity $C_B$ located very close to the Kemour mound. The anaglyphs were constructed from NAC images taken on 2015 July 18 and the indicated times are the mid-points of the two successive images forming a given anaglyph.

(vii) Whereas $C_A$ and $C_B$ are characterized by similar properties, $C_C$ stands out as somehow intermediate between these two cases and the surrounding terrains and as such, reveals the variety of small scale features on the nucleus.

(viii) The three cavities are probably native voids in a low density nucleus with inhomogeneities in the first few ten metres and probably down to a few hundred metres below its current surface as concluded by Vincent et al. (2015) from their study of pits.

(ix) As such, the icy cavities are the first potential subsurface access points detected on a cometary nucleus and their lifetimes suggest that they reveal pristine sub-surface icy layers and/or pockets.

## ACKNOWLEDGEMENTS

The construction of the stereo anaglyphs and the realization of the catalog and associated website were funded by the Centre National d'Etudes Spatiales. We thank C. Matonti for clarifying the origin of the Kemour mound. OSIRIS was built by a consortium of the Max-Planck-Institut für Sonnensystemforschung, Göttingen, Germany, CISAS-University of Padova, Italy, the Laboratoire d'Astrophysique de Marseille, France, the Instituto de Astrofísica de Andalucía, CSIC, Granada, Spain, the Scientific Support Office of the European Space Agency, Noordwijk, Netherlands, the Instituto Nacional de Técnica Aeroespacial, Madrid, Spain, the Universidad Politéchnica de Madrid, Spain, the Department of Physics and Astronomy of Uppsala University, Sweden, and the Institut für Datentechnik und Kommunikationsnetze der Technischen Universität Braunschweig, Germany. The International ROSETTA Mission was a cooperative project between ESA, several European national space agencies, and NASA..

## DATA AVAILABILITY

The fully corrected and calibrated versions of all OSIRIS images are available on ESA's Planetary Science Archive. The anaglyphs are publicly accessible at https://rosetta-3dcomet.cnes.fr. Once published, the 67P-133MP shape model will be available on the above website.

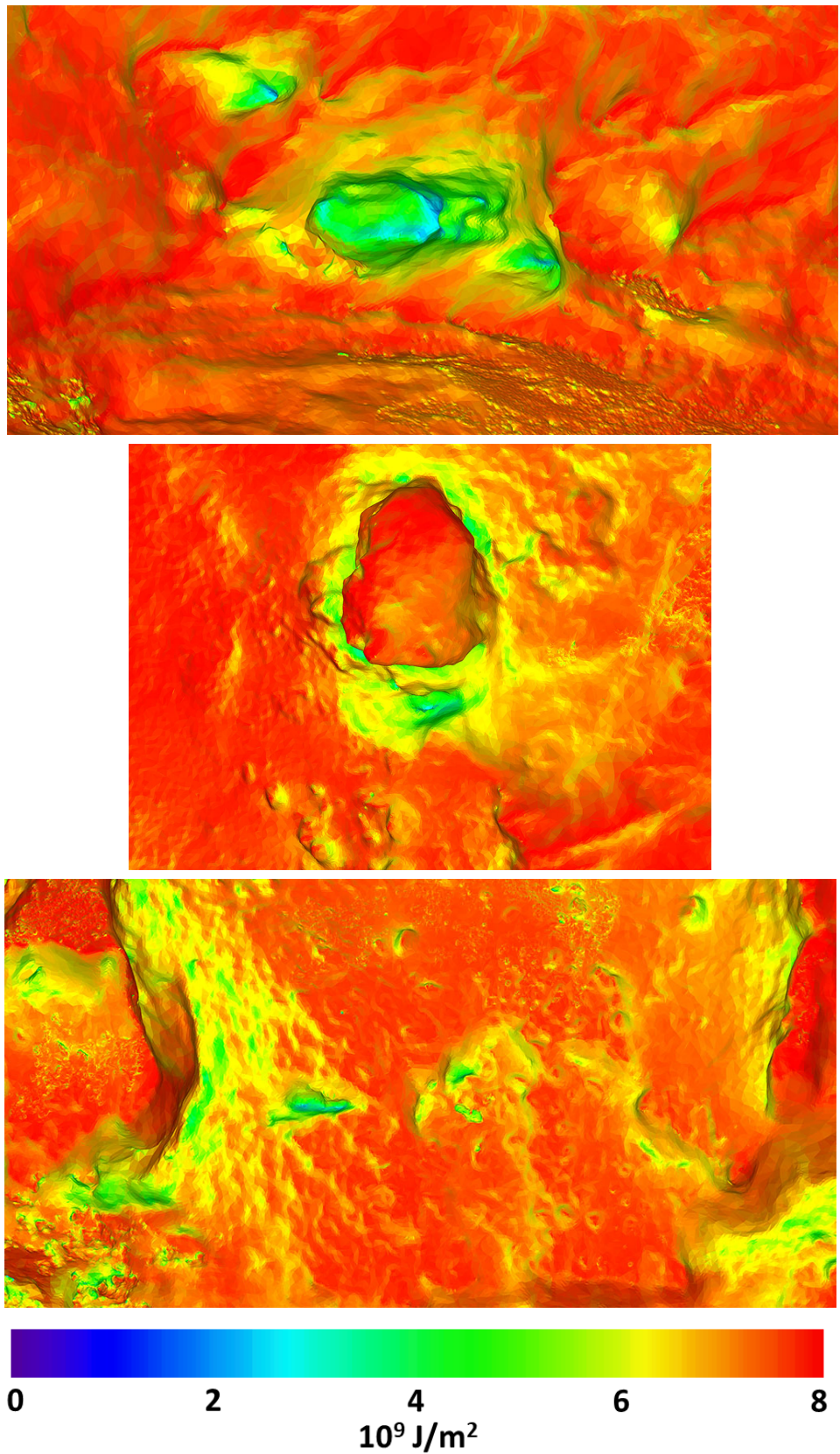

**Figure 23.** Maps of the accumulated energy received by cavities $C_A$ (upper panel), $C_B$ (middle panel), $C_C$ (lower panel), and their surrounding terrains expressed in units of $10^9\,\mathrm{J\,m^{-2}}$.

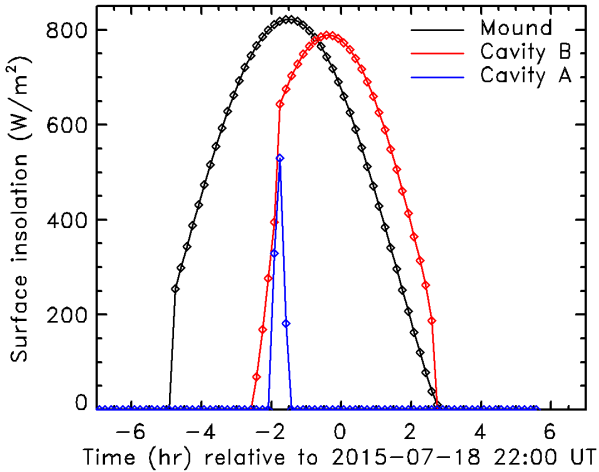

**Figure 24.** Temporal evolution of the insolation of cavities $C_A$ and $C_B$ and of the top of the Kemour mound.

## APPENDIX A:  TEXTURE MAP

A realistic rendering of the cometary surface was obtained by applying the standard technique of UV mapping in which the texture is derived from the original images themselves. The two-dimensional UV texture map – where U and V stand for the two coordinates of the map – was constructed on a grid of $16{,}384 \times 16{,}384$ pixels to ensure an adequate spatial scale on the nucleus, typically 40 cm per pixel. At a given pixel, that is a surface element on the nucleus, the texture was calculated as the mean of the radiance from all images which contain this element. This first UV map appears too inhomogeneous as a result of the diversity of illumination and observing conditions. The distribution of radiance was balanced by darkening the bright parts and lightening the somber ones. We finally obtained a quasi Gaussian distribution that extends over a realistic shade of gray. This final UV texture map illustrated in Figure A1 was applied to the surface of the shape model of the nucleus.

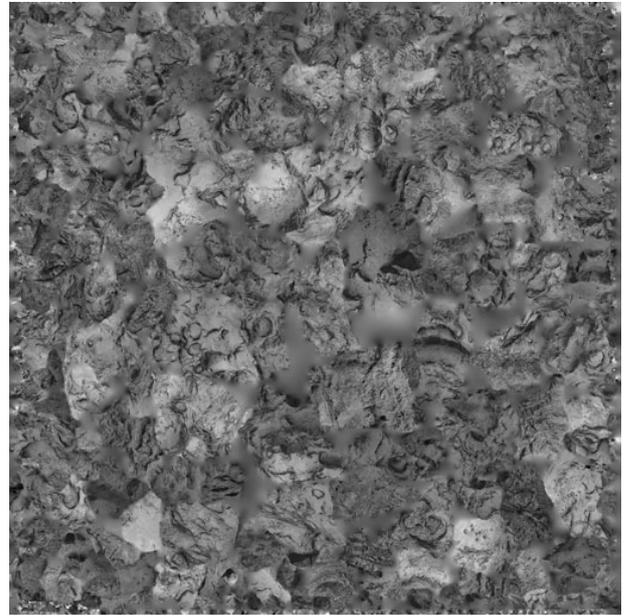

**Figure A1.** Final UV texture map used to create the image and anaglyph displayed in Figure 8.

## APPENDIX B:  ON THE CORRECTION OF THE RADIANCE FACTOR

The radiance factor $I/F$ is a standard quantity appropriate to the disk-resolved photometry of solar system airless bodies and it is defined by the following equation:

$$I/F = \frac{\pi \, I(i, e, \alpha, \lambda)}{F_\odot(\lambda)} \tag{B1}$$

where $I(i, e, \alpha, \lambda)$ stands for the measured spectral radiance of the body and $F_\odot(\lambda)$ for the solar spectral irradiance at the heliocentric distance of the body, all at wavelength $\lambda$. So, like the spectral radiance $I$, the radiance factor is a function of i) the illumination and viewing geometry via the incidence and emergence angles $i$ and $e$ and the phase angle $\alpha$ of the observations and ii) the scattering properties of the surface of the body determined by many parameters, prominently mineralogical composition and typology of the terrain (e.g. regolith versus consolidated material), all depending upon $\lambda$. In order to disentangle the extrinsic geometric conditions and the intrinsic physical characteristics of the surface, the standard approach consists in relying on photometric models. Apart from the wavelength and solar phase angle of the observations, they require the incidence and emergence angles for each pixel which are calculated from a 3D shape model of the body of adequate spatial resolution. Various formalisms are available and were thoroughly discussed by La Forgia et al. (2015) in the framework of their photometric analysis of the Agilkia region of 67P. We present below a short description emphasizing the most important aspects.

The Hapke model (Hapke (1981) and subsequent modifications) relies on an empirical formulation with multiple ad-hoc parameters specifying the different scattering processes at work (e.g. opposition surge, particle phase function) to express the bidirectional reflectance (BDR) of a surface. Apart from the fact that the Hapke parameters are difficult to interpret in term of physical quantities – thus favoring a direct comparison of BDR measurements retrieved from observation with laboratory ones (e.g. Spjuth et al. (2012)) – the model presents well-known shortcomings. Comparisons with laboratory measurements indeed revealed that individual photometric parameters could not be uniquely determined either in an absolute or relative sense (Shepard & Helfenstein 2007). There is in particular an ambiguity between the roughness parameter and the asymmetry factor.

Simpler and widely used photometric models implement a phenomenological approach where the radiance is simply expressed by the product of two functions, the disk function $D(i, e)$ of the incidence and emergence angles at a constant phase angle and the phase function depending solely upon the phase angle $\alpha$. When the phase angle coverage of the observations is small, the second function is disregarded and the model is thus restricted to the disk function. Consequently, the radiance factor is corrected according to:

$$I/F_{cor}(\alpha, \lambda) = \frac{\pi I(i, e, \alpha, \lambda)}{F(\lambda) D(i, e)}. \tag{B2}$$

Popular disk functions for small airless bodies are the Lommel–Seeliger law and the Akimov function.

Considering the specific case of the nucleus of 67P, Fornasier et al. (2015) implemented the 1993 version of the Hapke model (Hapke 1993) to perform its disk-averaged photometry, a version that does not include the effect of porosity later introduced by Hapke (2008). The work by Fornasier et al. (2015) remains the only application of an Hapke model to 67P so far. As a historical note, available disk-resolved photometric analyses of a cometary nucleus with a Hapke model remain those carried out by Li and co-workers in the case of 9P/Tempel 1 (Li et al. 2007a), 19P/Borrelly (Li et al. 2007b), 81P/Wild 2 (Li et al. 2009), and 103P/Hartley 2 (Li et al. 2013).

For performing the disk-resolved photometry of 67P, Fornasier et al. (2015) turned to the simple approach of the Lommel–Seeliger law:

$$D(i, e) = \frac{2 \cos(i)}{\cos(i) + \cos(e)} \tag{B3}$$

and this became the standard in several subsequent photometric articles, for instance Deshapriya et al. (2016), Feller et al. (2016), and Oklay et al. (2016). In support of this choice, Fornasier et al. (2015) argued that "the Lommel-Seeliger law comes from the radiative transfer theory when considering a single-scattering particulate surface". As pointed out by Hapke (2008), "it is well known that the bidirectional reflectance of a particulate medium such as a planetary regolith depends on the porosity, in contrast to predictions of models based on the equation of radiative transfer which, as usually formulated, is independent of porosity". Various considerations suggested that a probable reason for the failure of models based on the radiative transfer equation lies in the assumption that the medium is continuous. This led him to adapt the equation of radiative transfer to a medium made of discrete layers of particles and finally propose an improved new version of his model that then includes the effect of porosity. As far as we know, the only application of this version was performed by Spjuth et al. (2012) in the case of asteroid (2867) Steins and they found that its surface is indeed highly porous ($\approx 84\,\%$).

It becomes clear that the Lommel-Seeliger law was widely implemented because of its simplicity, but that it has no serious justification on theoretical grounds. In addition, it is barely conceivable that a unique law could apply to surfaces of very different geomorphology encompassing consolidated material, diamicton deposit, fine grain deposit, and outcropping layered terrain to mention a few examples observed on the nucleus of 67P. As a noteworthy exception, La Forgia et al. (2015) used instead the parameter-free version of the Akimov function whose expression is more elaborated than the Lommel-Seeliger law as it involves the phase angle and the photometric latitude and longitude, both depending upon the incidence, emergence, and phase angles (see their Equations 5, 6 and 7). On the one hand, they validated this function by the remarkable agreement between its resulting reflectance and that obtained by the Hapke solution obtained by Fornasier et al. (2015). On the other hand, they conceded that they had to assume that the disk-averaged Hapke parameters hold for the whole Agilkia region under study, a questionable assumption in view of the variety of terrains illustrated by their fig. 5.

We point out another caveat in the case of photometric analysis at high spatial resolution since the sampling of the 3D SHAP7 model may not match the pixel scale of the involved images. Indeed, the optimum number of facets of the shape model should be such that the image pixel corresponds to one or just a few facets and this condition may not be satisfied. This points to the superiority of the method pioneered by Spjuth et al. (2012) which operates in the space of the facets, whereas all past analysis have always been performed in the space of the image pixels although they are not intrinsic to the surface of the body. This imposes that the sampling of the shape model be appropriate with respect to the pixel scale of the images, but has the decisive advantage of automatically tracking the same local surface element on a series of images.

This paper has been typeset from a TeX/LaTeX file prepared by the author.